\documentclass[aps,twocolumn,showpacs,amsmath,amssymb,prx,superscriptaddress,floatfix,longbibliography]{revtex4-1}
\usepackage{ braket}
\usepackage{amsmath}
\usepackage{amssymb}
\usepackage{mathtools}
\usepackage{graphicx}
\usepackage{dcolumn}
\usepackage{bm}
\usepackage{multirow}
\usepackage{leftidx}
\usepackage{color}
\usepackage{qcircuit}
\usepackage{ifthen}

\usepackage{amsfonts}
\usepackage[german,english]{babel}

\newcommand \be{\begin{equation}}
\newcommand \ee{\end{equation}}
\newcommand \bea{\begin{eqnarray}}
\newcommand \eea{\end{eqnarray}}
\newcommand \bse{\begin{subequations}}
\newcommand \ese{\end{subequations}}

\newcommand {\rsub}[1]{\textcolor{black}{#1}}

\newboolean{ShowComments}
\setboolean{ShowComments}{false}  

\definecolor{mscolor}{rgb}{0,0.5,0.5}
\definecolor{tgcolor}{rgb}{0.5,0,0.5}

\definecolor{rkccolor}{rgb}{0.5,0.5,0}

\begin{document}
\title{
Mid-circuit measurements on a single species neutral alkali atom quantum processor} 
\author{T. M. Graham}
\affiliation{Department of Physics, University of Wisconsin-Madison,
Madison, Wisconsin 53706}
\author{L. Phuttitarn}
\affiliation{Department of Physics, University of Wisconsin-Madison,
Madison, Wisconsin 53706}
\author{R. Chinnarasu}
\affiliation{Department of Physics, University of Wisconsin-Madison,
Madison, Wisconsin 53706}
\author{Y. Song}
\affiliation{Department of Physics, University of Wisconsin-Madison,
Madison, Wisconsin 53706}
\affiliation{current address: Korea Research Institute of Standards and Science, Daejeon 34113, Republic of Korea}
\author{C. Poole}
\affiliation{Department of Physics, University of Wisconsin-Madison,
Madison, Wisconsin 53706}
\author{K. Jooya}
\affiliation{Department of Physics, University of Wisconsin-Madison,
Madison, Wisconsin 53706}
\author{J. Scott}
\affiliation{Department of Physics, University of Wisconsin-Madison,
Madison, Wisconsin 53706}
\author{A. Scott}
\affiliation{Department of Physics, University of Wisconsin-Madison,
Madison, Wisconsin 53706}
\affiliation{current address: Department of Computer Science, University of Colorado, Boulder, Colorado, }
\author{P. Eichler}
\affiliation{Department of Physics, University of Wisconsin-Madison,
Madison, Wisconsin 53706}
\author{M. Saffman}
\affiliation{Department of Physics, University of Wisconsin-Madison,
Madison, Wisconsin 53706}
\affiliation{Infleqtion Inc., Madison, Wisconsin, 53703}

\date{\today}

\begin{abstract}
We demonstrate mid-circuit measurements in a  neutral atom array by shelving data qubits in protected hyperfine-Zeeman sub-states while non-destructively measuring an ancilla qubit. Measurement fidelity was enhanced using   microwave repumping of the ancilla during the measurement. The coherence of the shelved data qubits was extended during the ancilla readout with dynamical decoupling pulses, after which the data qubits are returned to $m_f=0$ computational basis states.  We demonstrate that the quantum state of the data qubits is well preserved up to a constant phase shift  with  a state preparation and measurement (SPAM) corrected process fidelity of ${\mathcal F}= 97.0(5)\%$. 
The measurement fidelity on the ancilla qubit after correction for state preparation errors is ${\mathcal F}=94.9(8)\%$ and ${\mathcal F}=95.3(1.1)\%$ for $\ket{0}$ and $\ket{1}$ qubit states, respectively.
We discuss extending this technique to repetitive quantum error correction using quadrupole recooling and microwave-based quantum state resetting. 
\end{abstract}

\maketitle

\section{Introduction}

Scalable quantum computation relies on the coherent, unitary evolution of a large number of qubits. Available qubit technologies have limited coherence times and suffer from errors in gate operations which restricts the usable length of quantum circuits\cite{Preskill2018}. To circumvent these limitations, the leading approach to quantum error correction combines repetitive mid-circuit projective measurements of ancilla qubits for error syndrome extraction, with corrective operations (or frame changes) that are conditioned on the results of measurement outcomes. Repetitive mid-circuit measurements have been demonstrated on quantum processors that use 
trapped ion\cite{Schindler2011,Negnevitsky2018,RyanAnderson2021,Egan2021}, 
superconducting\cite{Kelly2015,Ofek2016,LHu2019,Andersen2020,Bultink2020,ZChen2021,Acharya2023}, 
or spin\cite{Cramer2016} 
qubits. Mid-circuit measurements are also valuable for reducing circuit complexity and increasing the fidelity of a targeted output state, without invoking the full machinery of error correction.  An early example of the utility of mid-circuit measurements was provided by the semi-classical quantum Fourier transform\cite{Griffiths1996,Chiaverini2005}. Such semi-classical techniques use the information provided by mid-circuit measurements to prepare a target state without two-qubit gate operations. More recently, several works have shown how mid-circuit measurements or qubit reset and reuse may be used to extend qubit coherence\cite{Singh2023}, improve the performance of noisy hardware for state preparation\cite{Rattew2021,TCLu2022} or increase the efficacy of noise mitigation techniques\cite{Czarnik2021}.

Fast, reliable mid-circuit measurements in neutral atom quantum computing systems face several challenges. Quantum state measurements of qubits encoded in atomic hyperfine-Zeeman states are generally based on detection of scattered photons from the state that is coupled to imaging light, whereas the other qubit state does not couple, or only couples very weakly, to the light used for imaging\cite{Saffman2005a}. However, during readout, off-resonant Raman transitions cause depumping to states in a far-off-resonant hyperfine manifold that is dark to the readout light, so a repumping laser is needed. The standard protocol for performing state-selective measurements with alkali atom qubits is to use a resonant push-out beam to remove atoms in one of the two computational basis states. Atoms in the remaining computational basis state are then made to fluoresce using near-resonant light. The fluorescence is imaged onto a sensitive camera where it is detected and spatially analyzed to determine the array occupancy. This measurement protocol is incompatible with mid-circuit measurements for two main reasons. First, it is global; a mid-circuit measurement requires only selected sites to be measured without affecting the states of non-selected sites. Second, it is destructive; a qubit measured during a mid-circuit measurement should be reusable in the remainder of the quantum circuit. If the imaging beam is focused to selectively readout an ancilla qubit, then it is possible to achieve a high fidelity, site-selective state measurement, but neighboring data qubits suffer from the fact that they may absorb a scattered photon, leading to a change in the quantum state. 

There are several strategies to address these issues including electromagnetically induced transparency \cite{Giraldo2022,Saglam2023}, atom transport to isolated imaging regions\cite{Bluvstein2022}, the use of multiple atomic species\cite{Beterov2015,Singh2023,JTZhang2022}, or isotopes\cite{YZeng2017,CSheng2022}, and the use of atoms\cite{YWu2022}, or ions\cite{Postler2022}, with a more complex internal structure that allows measurements to be performed using wavelengths that do not interact with atoms in computational basis states. Actual demonstrations of neutral atom mid-circuit measurements have used atom transport into a near-concentric readout cavity\cite{Deist2022}, and a two-species array with one of the species acting as  spectator qubits that provided in-situ noise characterization to enable coherent correction of errors on data qubits\cite{Singh2023}.

We present here the first scalable mid-circuit measurements on a single alkali species neutral atom processor compatible with global, parallel ancilla readout. Measurements are implemented in an array of Cs atom qubits and characterized as regards both the measurement fidelity of the ancilla qubit and how the mid-circuit measurement affects the data qubits. 
As described in Sec. \ref{sec.experiment} the approach is  based on  shelving data qubits in hyperfine levels that are only weakly coupled to the light used for state measurement, while the ancilla qubit is measured using previously demonstrated non-destructive state-selective readout techniques \cite{Kwon2017,Martinez-Dorantes2017}. The data qubits are restored to the qubit basis upon completion of the measurement.  The obtained results for ancilla qubit measurements and process fidelity of the quantum information encoded in a  plaquette of data qubits are presented in Sec. \ref{sec.results}.
We conclude in Sec. \ref{sec.outlook} with an outlook for further improvements and applications to quantum state preparation and measurement based quantum computation.  \rsub{Technical details of the apparatus as well as detailed atomic calculations that clarify the ultimate performance limits of the approach used are given in Appendices A-F. }

\section{Experimental approach}
\label{sec.experiment} 

The experimental apparatus has been described in detail in \cite{Graham2022}; however, some modifications were made to facilitate mid-circuit measurements. Cs atoms were laser cooled and stochastically loaded into a $3\times 3$ two-dimensional array of red-detuned, 1064-nm wavelength tweezers which had a beam waist of $1.2~\mu$m and a trap depth of approximately $0.4~$mK. The tweezer array was created by diffracting a Gaussian input beam with a two-dimensional acousto-optic deflector (AOD). Each of the deflector's axes was driven with three different frequency tones provided by a Quantum Machines OPX device. The amplitudes of the frequency tones were adjusted for equal trap depths by  balancing the trap light induced Stark shifts \footnote{Ramsey microwave spectroscopy was used to determine the precise clock state resonant frequency, which was then used as a feedback signal to adjust the trap frequency tone amplitudes}.

\begin{figure}[!t]
\includegraphics[width=3.3in]{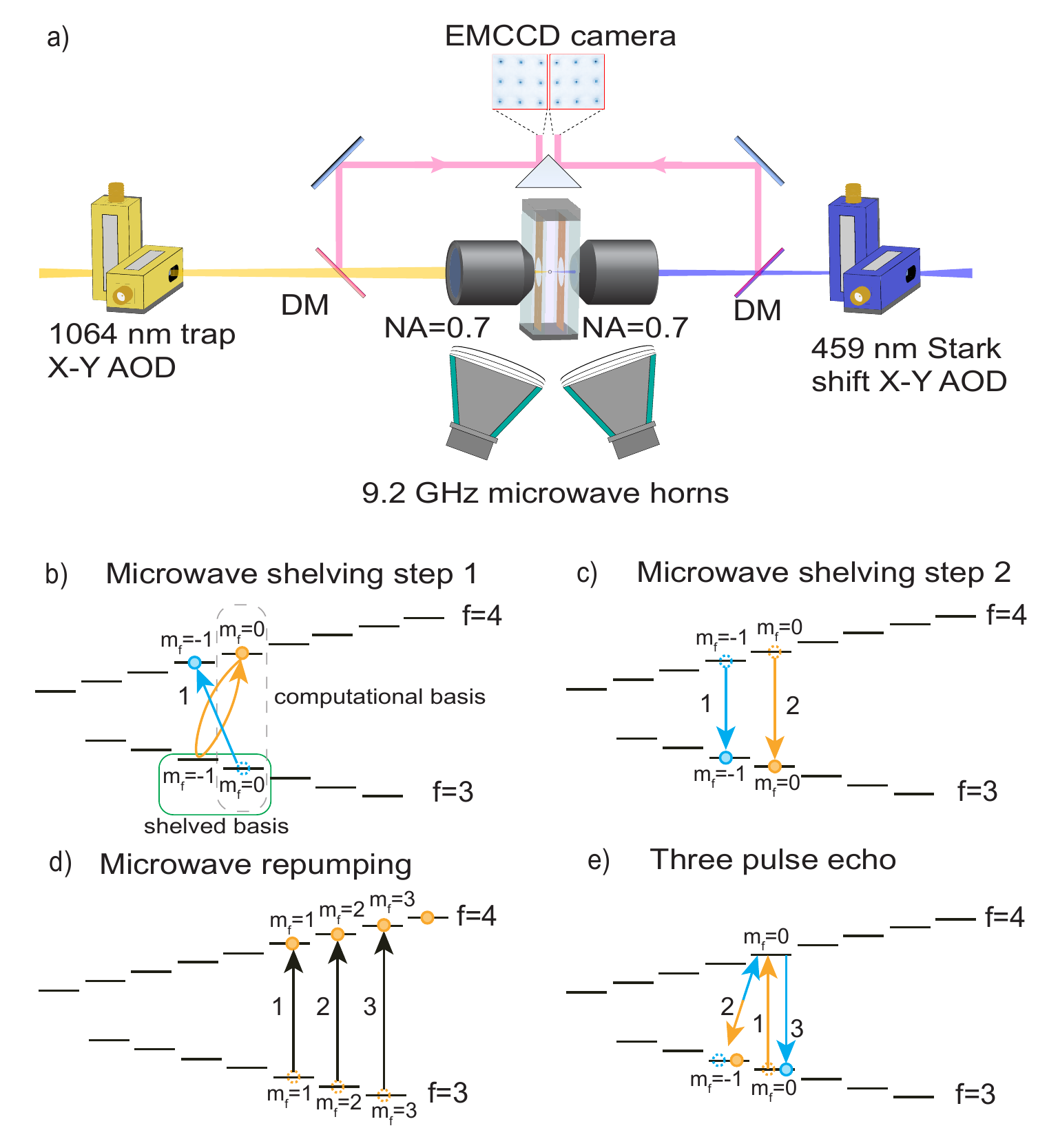}
\caption{
\label{fig.1}
a) Experimental layout for mid-circuit measurement. Atoms were trapped and cooled into a $3\times3$ 1064-nm tweezer array.   The microwave horns are each attached to 40 W amplifiers that are driven by 9.2 GHz signals. A 459-nm beam provided site-selective Stark shifts on targeted sites. During detection, near-resonant 852-nm light was used to induce atomic fluorescence. The fluorescence was imaged onto an electron-multiplying CCD camera. b) Data qubits prepared in the computational basis ($f=3, m_f=0$ and $f=4, m_f=0$, enclosed by the grey dashed line) were shelved in the $m_f=0$ and $m_f=-1$ Zeeman states of the $f=3$ hyperfine manifold (enclosed by the solid green line) in a two-step process.  In the first step, a microwave pulse  drove transitions from $\ket{4,0}\rightarrow \ket{3,-1}$ and $\ket{3,0}\rightarrow \ket{4,-1}$. A relative phase shift between the two microwave horns provided control of the microwave field polarization such that the former transition undergoes a $2\pi$ rotation while the latter undergoes a $\pi$ rotation. The atoms were thereby stored in the $f=4$ manifold. Note that during this pulse, the ancilla qubit was shifted out of resonance using the 459-nm beam to provide a site-selective Stark shift\cite{Xia2015}. c) In the second step the atoms were transferred from $f=4$  to the corresponding Zeeman states in the $f=3$ hyperfine manifold using two composite microwave pulses. d) During mid-circuit measurement of the ancilla, excited state hyperfine mixing can cause atoms in the bright $f=4$ state to leak into $f=3$. Three microwave pulses were periodically employed to transfer population back up to $f=4$ to allow fluorescence measurements to resume. e) During the mid-circuit measurement, the data qubits experience dephasing due to magnetic fluctuations. Periodic echoing using three composite microwave pulses was used to mitigate the dephasing.     
}
\end{figure}

\begin{table}[!t]
\caption{Readout parameters. $\gamma=2\pi\times 5.2\rm$ MHz is the decay rate of the Cs $6p_{3/2}$ state and $I_{\rm sat}= 11.0~\rm W/m^2$ is the saturation intensity of the $6s_{1/2}\ket{4,4}\rightarrow 6p_{3/2}\ket{5,5}$ transition.  }\label{Table.1}
\centering
\begin{tabular}{|c|c|c| }

 \hline
     & Occupation readout  & Mid-circuit readout\\
\hline

Detuning & $-9 \gamma$ & $-2 \gamma$   \\
Intensity & $5 I_{\rm sat}$ & $3 I_{\rm sat}$ \\
Trap depth$/k_{\rm B}$  &  0.85 mK & 1.8 mK \\
Readout time & 30 ms & 4 ms\\

\hline
\end{tabular}
\end{table}
Measurement of the initial atom occupancy was performed using two pairs of counter-propagating $\sigma_+/\sigma_-$ polarized beams. The four beams lay in \rsub{the plane of the atom array.} The beams were red-detuned from the $6s_{1/2},f=4 \leftrightarrow 6p_{3/2},f=5$ transition. During the measurement, atoms were occasionally depumped into the $6s_{1/2},f=3$ state which is dark to the readout light. Light tuned to the $6s_{1/2},f=3 \leftrightarrow 6p_{3/2},f=4$ transition  was added to each of the four readout beams to repump atoms back to the $f=4$ state. Typical readout parameters are summarized in Table \ref{Table.1}. To avoid standing waves, we used fiber acousto-optic modulators to apply 1-4 kHz frequency shifts to each readout beam. We also adiabatically ramped the trap depth to 0.85 mK to reduce loss from heating during the readout. The readout light was chopped out-of-phase with the trapping light using a $50\%$ duty cycle and a chopping frequency of $1.42$ MHz. Chopping the trap prevented the atoms from experiencing trap-induced Stark shifts while being illuminated by the readout beams. The atom occupancy was determined from a $30$ ms camera exposure.

After measuring the trap occupancy, the atoms were recooled using a combination of red-detuned polarization gradient cooling on the D2 line and blue-detuned lambda grey molasses using the D1 line \cite{YFHsiao2018} cooling atoms to a temperature of $7~\mu$K. The atoms were then optically pumped into the $\ket{f=4,m_f=4}$ (henceforth denoted $\ket{4,4}$) stretched state using $\sigma_+$ polarized  895-nm light. Photon scattering from optical pumping heated the atoms up to a temperature of $10~\mu$K. We then used a series of CORPSE pulses, composite microwave pulses which are robust against off-resonant errors \cite{Cummins2003}, to transfer the atoms to the $\ket{3,0}$ state ($\ket{0}$ in the qubit computational  basis) (see Appendix \ref{sec.microwave}). After these microwave operations, all atoms were initialized to $\ket{3,0}$. \rsub{In the rest of the paper we refer to the qubit basis states as $\ket{0}=\ket{3,0}$  and $\ket{1}=\ket{4,0}$. }

\subsection{Mid-circuit measurement}

The mid-circuit measurement procedure is composed of 5 steps: 
1) ramp up the trap power on the ancilla atom to be measured, 
2) shelve data qubits (qubits that are not being measured) into the $f=3$ hyperfine manifold, 
3) perform state-selective readout of atoms in the $\ket{0}$ state, 
4) unshelve the data qubits to restore the quantum states after the readout, and 
5) ramp the ancilla site trap back to the original level. 
\rsub{Note that because of the swapping of the $\ket{0}$ and $\ket{1}$ states of all qubits, including the ancillas, during the measurement process to be explained below,  it is ancillas that are initially in the $\ket{0}$ state that result in a bright readout signal.}   
During this process it is important that all data qubits maintain good phase coherence between $\ket{0}$ and $\ket{1}$ states to prevent the degradation of the qubits. Since the measured qubit is projected into one of the two computational qubit basis states, the coherence between these two states has no importance for the ancilla. However, for the mid-circuit measurement to be compatible with repetitive error correction schemes, the ancilla must be able to be reinitialized back to $\ket{0}$ and recooled back to its initial temperature. This qubit resetting was not performed in this initial mid-circuit measurement demonstration.  Appendix \ref{sec.qubit_reset} presents a path to implementing qubit reset in future experiments. Below, we detail how we implement the mid-circuit measurement technique in the atomic qubit array. 

\subsubsection{Trap Ramping}

To prepare for mid-circuit measurements, we first increased the trap depth of the ancilla site. The state selective measurement technique used for the mid-circuit measurement only cools the atoms in one dimension using counter-propagating $\sigma_+/\sigma_+$ polarized beams for the readout. While this polarization orientation minimizes hyperfine depumping, it  prevents polarization gradient cooling during the readout. Only 1D Doppler cooling is present during the readout and  the ancilla atom is heated in the other two dimensions.  A deeper trap depth allows more photons to be scattered from the ancilla atom before it is heated out of the trap. However, raising the trap depths of the data qubits was not desirable as it heated the data qubits, resulting in faster dephasing and required more total trap power. Instead, we increased the central frequency tone amplitude on each axis of the trap while simultaneously lowering the other two frequency tones. This increased power in the central trap site of the $3\times 3$ array while each of its nearest neighbor sites decreased, further cooling the data qubits \footnote{During the trap ramping, the depth of the four corner traps is lowered, and are thus not considered for experimental demonstration. In future applications, these sites can be used as a reservoir of atoms for atomic rearrangement.}. We use interpolation between the start and end points to smoothly transition between the two settings using a Blackman ramp profile over $200~\mu$s.  

\subsubsection{Data qubit shelving}

To perform a mid-circuit measurement, all data qubits needed to be shelved in the $f=3$ hyperfine manifold to avoid interacting with the readout light; shelved qubits are protected by a $9.2$ GHz detuning. This task is complicated by the fact that the transitions $\ket{3,0} \leftrightarrow \ket{4,-1}$ and $\ket{4, 0} \leftrightarrow \ket{3, -1}$ have nearly the same resonant frequency at the $10.2$ G magnetic bias field that is used in this demonstration. Simply driving the transition to move population from $\ket{4,0} \rightarrow \ket{3,-1}$ in an effort to shelve the $\ket{1}$ qubit state in the $f=3$ manifold, would transfer  population from $\ket{0}=\ket{3,0}$  to $\ket{4,-1}$. To circumvent this problem, we used two microwave horns driven at the same frequency at different orientations. By adjusting the relative driving phases of these horns we controlled the microwave polarization, as described in \cite{Smith2013} (for alternative shelving methods, see Appendix \ref{sec.shelving}). Each horn was driven with a phase controlled 9.2 GHz signal provided by a 40 W amplifier (for microwave system details, see Appendix \ref{sec.microwave}). Given physical space constraints and incomplete microwave polarization control, we found that it was optimal to perform the shelving sequence in two steps. In the first step, we shelve in  $f=4$ by adjusting the microwave polarization such that the $\ket{3,0} \rightarrow \ket{4,-1}$ transition experiences a $\pi$ rotation while $\ket{4,0} \rightarrow \ket{3,-1}$  experiences a $2\pi$ rotation. After this microwave pulse, state $\ket{0}$ was transferred to  $\ket{4,-1}$, and state  $\ket{1}$ remained encoded in $\ket{4,0}$. We then used two composite CORPSE pulses to transition $\ket{4,0} \rightarrow \ket{3, 0}$ and $\ket{4,-1} \rightarrow \ket{3, -1}$.

During the shelving sequence, the ancilla atom must be prevented from being shelved, or it will not be detected. We used  459-nm, $\pi$ polarized light that is red-detuned from the $6s_{1/2},f=4 \leftrightarrow 7p_{1/2}$ center of mass by 24 GHz to shift the atom out of resonance with the first half of the two step shelving process, leaving the ancilla qubit encoded in the clock state basis \cite{Xia2015}.  The CORPSE pulses then swap population between the two clock states. At this point, we detected the population in $\ket{4, 0}$ using readout light. 

\subsubsection{State-selective qubit measurement}

To measure the state of the ancilla qubit, we require population in the $\ket{1}$ state to fluoresce without depumping to the $f=3$ hyperfine manifold. To prevent depumping, these measurements were performed with one pair of counter-propagating beams with $\sigma_+/\sigma_+$ polarization with respect to a $10.2$ G bias magnetic field. When illuminated, an atom occupying any state in the $f=4$ manifold was pumped to the $\ket{4,4}$ stretched state and then cycled on the $6s_{1/2}\ket{4,4} \leftrightarrow  6p_{3/2} \ket{5,5}$ transition \cite{Kwon2017,Martinez-Dorantes2017}.  With this readout beam and magnetic bias geometry, Raman transitions to the $6s_{1/2}$, $f=3$ hyperfine manifold are forbidden by selection rules. However, such methods need to be adapted to be compatible with the mid-circuit measurement protocol. Tensor shifts from the trap cause excited state mixing that result in depumping the ancilla to the $f=3$ manifold. While this problem can be mitigated by chopping the trap and readout light out of phase with each other, as is done in the initial occupancy measurement, the chopping causes either heating of the data atoms at lower chopping frequencies ($\leq 1.3$ MHz) or Zeeman state mixing at higher chopping frequencies ($\geq 1.3$ MHz). These limitations result in the ancilla atom occasionally being depumped into the $f=3$ manifold during readout. However, we found that periodically employing microwave pulses to coherently transfer population from $m_f>0$ Zeeman states in $f=3$ to the corresponding Zeeman state in $f=4$ mitigated this depumping problem (see Fig. \ref{fig.1}d). 

During the mid-circuit measurement, the data qubits are shelved in a magnetically sensitive state, and have a relatively short $T_2^*=3.2$ ms coherence time due to magnetic field and vector Stark shift fluctuations. To accommodate the limited $T_2^*$ time, we applied dynamical decoupling pulses to the data qubits during the readout sequence. We did not have a way of directly driving the $\ket{3,0}\leftrightarrow\ket{3,-1}$ transition, so we periodically paused the readout and performed a three pulse echo sequence (see Fig. \ref{fig.1}e). This sequence is comprised of a $\pi$-pulse to transfer population from $\ket{3,0}$ to $\ket{4,0}$, followed by a $\pi$-pulse to exchange population between $\ket{4,0}$ and $\ket{3,-1}$. A final clock state $\pi$-pulse transfers population from $\ket{4,0}$ back to $\ket{3,0}$. One complication arose in the echoing due to the differences between the trap depths of the ancilla qubit and the data qubits. To accurately determine the state of the ancilla qubit, atoms occupying the state $\ket{3,0}$ at the start of the readout must remain dark; however, the trap Stark shift changed the energy levels of the ancilla qubit such that the echoes were detuned. It was necessary to use CORPSE pulses during the echo to prevent the dark state of the ancilla qubit from leaking into the $f=4$ manifold. In addition, 459-nm light was used to shift the second composite echo pulse closer to resonance. During the 4 ms mid-circuit measurement, we found that 8 Hahn echo pulses were optimal to retain qubit coherence and minimize error due to imperfect microwave rotations.

To minimize dephasing on the data qubits, it was necessary to measure the ancilla atom quickly. This requirement suggested that we should use readout light with small detuning from resonance. However, since the trap is left at a constant intensity during the readout, an ancilla atom in the bright state experiences heating due to dipole force fluctuations (DFF) in addition to heating due to photon recoil (see \cite{Martinez-Dorantes2018} for an extensive treatment of these effects).  To reduce DFF heating, we used a detuning of $-2\gamma$ from the shifted resonance (shifted $+49$ MHz due to the trap Stark shift and Zeeman effect of the bias field) to scatter enough photons to distinguish the bright and dark states of the ancilla atom in 4 ms. This detuning choice balanced the effects of DFF heating and the dephasing of the data qubits. See Table \ref{Table.1} for a summary of readout parameters.

After the mid-circuit measurement, the ancilla atom's dark state was reset to its original state by performing the shelving sequence in reverse order. However, the bright state was significantly heated during the readout and was no longer suitable to use as a qubit. In future work, we will use a quadrupole transition to recool ancilla atoms in the bright state, so they can be restored to the computational basis and reused in the remainder of the calculation (see Appendix  \ref{sec.qubit_reset} for details).

\section{Results}
\label{sec.results}

\begin{figure}[!t]
\includegraphics[width=3.5in]{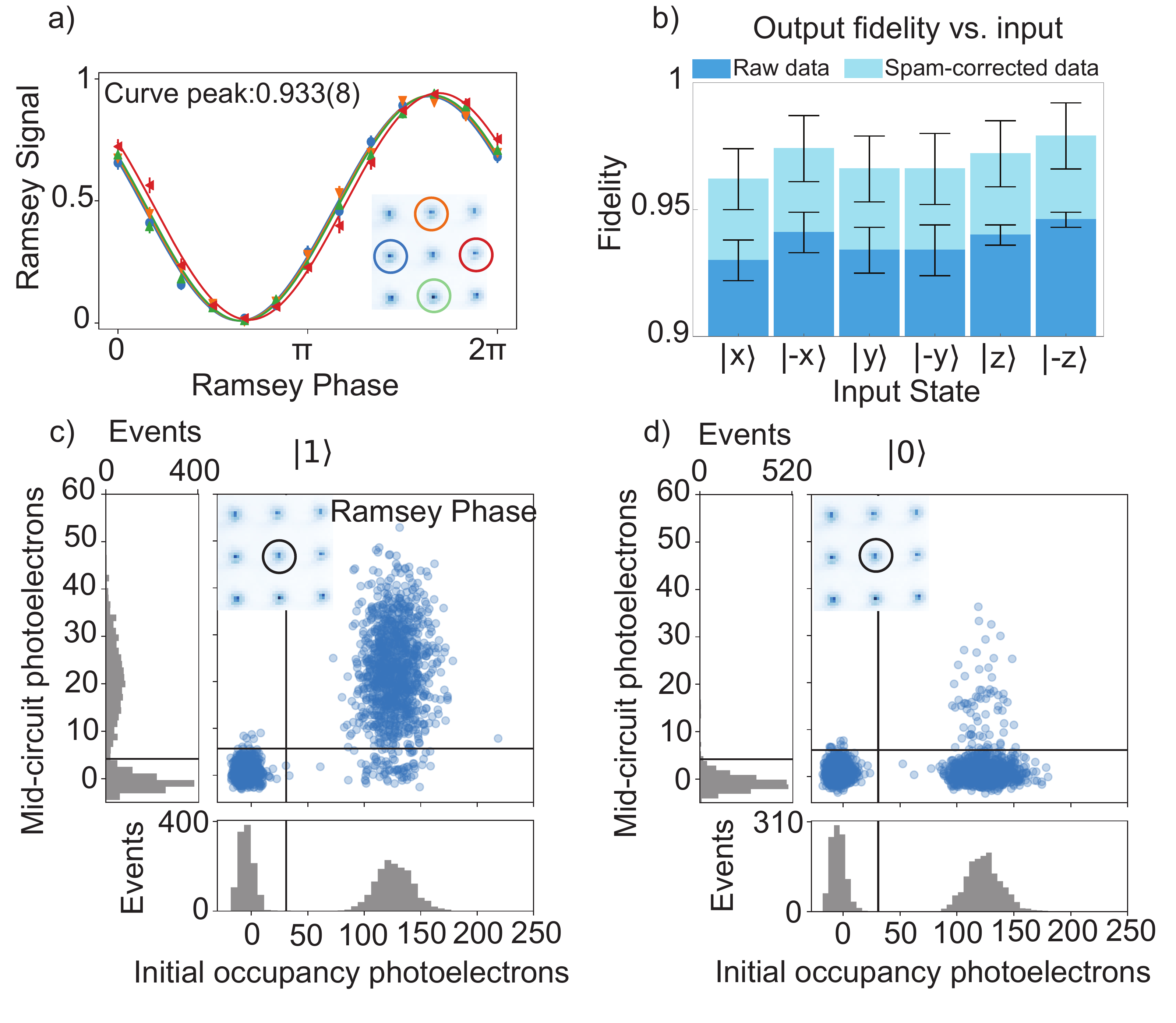}
\caption{
\label{fig.Data} Mid-circuit measurement results. 
a) To show that the data qubits retain coherence during the mid-circuit measurement, we prepared the atoms in $\frac{1}{\sqrt{2}}\left( \ket{0}+\ket{1} \right)$ using a $\pi/2$-microwave pulse and then performed a mid-circuit measurement on the ancilla.  After the measurement, we applied another $\pi/2$ rotation at a phase $\phi$ with respect to the original $\pi/2$-pulse and then measured the state of the data qubits. Ideally, a full contrast oscillation would be observed; however, imperfections in the microwave rotations limited the average peak-to-peak amplitude of the oscillation to $92.3(6)\%$.   b) Ideally, the mid-circuit measurement  acts as a simple phase rotation of the data qubits, which can be compensated through phase advances. We measured the average process fidelity of the mid-circuit measurement on the data qubits with a unitary phase advance. We accomplished this by characterizing the phase shift from the Ramsey curve shown in a). We prepared 6 mutually unbiased basis states and measured the fidelity with the expected state assuming the mid-circuit measurement acts as a unitary phase shift; the average process fidelity of $93.8(3)\%$ ($97.0(5)\%$ SPAM corrected) was obtained by averaging these 6 inputs.  Error bars represent 1 standard deviation from the mean. c,d) We characterized the fidelity of the ancilla measurement for a bright state input c) and a dark state input d). We observe a detection fidelity of $93.6(5)\%$ ($94.9(8)\%$ when compensating for state preparation errors) for a $\ket{0}$ input and $94.3(5)\%$ ($95.3(1.1)\%$ when compensating for state preparation errors) for a $\ket{1}$ input.   }
\end{figure}

There were two primary goals for the mid-circuit measurement demonstration. First, the ancilla qubit state must be accurately determined from the measurement; second, the state of the data qubits must not be changed by the measurement. To determine the ancilla qubit state, we imaged the atomic fluorescence onto an electron-multiplying CCD camera (EMCCD) and then applied standard region of interest and threshold techniques to determine if the atom was in a bright or dark state. To characterize the measurement fidelity, we sequentially performed measurements on each of the two computational basis state inputs (see Fig. \ref{fig.Data}). We obtained measurements with a $93.6(5)\%$ ($94.3(5)\%$) probability that a qubit prepared in $\ket{0}$ ($\ket{1}$)  was correctly measured. Some of this error was due to imperfect state preparation; when compensating for state preparation errors, we observed corrected fidelities of $94.9(8)\%$ ($95.3(1.1)\%$) for a qubit prepared in $\ket{0}$ ($\ket{1}$).

This measurement fidelity was limited by constraints imposed by the data qubits. The chief error sources for the dark state were imperfect state transfer during the echoing process, overlap of the bright and dark state histogram peaks, and incomplete shift-out of the microwave pulse which shelves the data qubits into the $f=4$ manifold. Bright state detection was similarly limited by histogram peak overlap and incomplete shift out of the microwave shelving pulse; in addition, it was limited by leakage into the shelved state basis as it was being pumped to the stretched state. Future improvements in fidelity may be achieved by using a higher power and further detuned shift-out laser, using quadrupole recooling during the readout to allow less detuned, faster readout, and using magnetic shielding and faster microwave sources to obtain higher fidelity microwave rotations.

In an ideal mid-circuit measurement, the data qubits would be completely unaffected by the measurement. The process of shelving and echoing the atoms during the measurement induced a phase shift on the data qubits; however, a phase advance can be compensated in quantum circuits in the circuit compilation or through a frame change. We  quantify the impact of the mid-circuit measurement on the data qubits by treating it as a quantum process and evaluating its fidelity relative to a phase gate with a calibrated phase shift. The phase shift was characterized by performing a mid-circuit measurement between two clock-state microwave $\pi/2$ rotations. By adjusting the phase of  the second pulse, we measured a Ramsey curve from which we extracted the phase (see Fig. \ref{fig.Data}).  We then determined the average process fidelity by measuring the fidelity of the inputs ($\ket{x}=\frac{\ket{0}+\ket{1}}{\sqrt{2}}$, $\ket{-x}=\frac{\ket{0}-\ket{1}}{\sqrt{2}}$, $\ket{y}=\frac{\ket{0}+i\ket{1}}{\sqrt{2}}$, $\ket{-y}=\frac{\ket{0}-i\ket{1}}{\sqrt{2}}$, $\ket{z}=\ket{0}$, $\ket{-z}=\ket{1}$) with target outputs calculated from using the extracted phase shift, and averaging the fidelities. This yielded a raw average process fidelity of $93.8(3)\%$. Compensating for state preparation and measurement errors, we obtained a fidelity of $97.0(5)\%$ (see Appendix \ref{sec.state_prep}). The process fidelity was limited by imperfect shelving and echoing and could be improved with faster microwave rotations. Faster readout of the ancilla state would also require fewer data qubit echoes preventing some of the microwave echo error accumulation.

\section{Outlook and discussion}
\label{sec.outlook}

This mid-circuit measurement demonstration is the first time single-species mid-circuit measurement of individual qubits has been demonstrated in a 2d array of neutral atom qubits and is a step towards error corrected and fault tolerant quantum computing. \rsub{Although the experimental results did not reach the fidelity needed for fault tolerant operation the analysis detailed in the Appendices supports reaching ancilla measurement and data errors approaching $10^{-3}$ with realistic experimental upgrades. }

\rsub{
Various other approaches to mid-circuit measurement have recently been explored experimentally. It is instructive to compare these approaches. Neutral atom mid-circuit readout approaches can be categorized into three groups. The first category to be demonstrated uses a combination of atomic motion and an isolated readout region and was demonstrated by Deist et al. \cite{Deist2022}. The data qubits are protected from the readout light as well as scattered light from the ancilla qubits by physical distance. The approach of Deist et al. used a strongly coupled in-vacuum cavity and a single photon counting module to achieve fast ancilla interrogation times of approximately $100 ~\rm \mu s$ with only about $1\%$ ancilla loss. This procedure achieved impressive results; however, even though the mid-circuit readout time was fast, since only one atom could be read out at a time, this technique would need to be significantly modified to be scaled to a large qubit array.} 

\rsub{
The second mid-circuit readout demonstration technique was performed by Singh et al.\cite{Singh2023} and used a dual-species atom array\cite{Singh2022}. In this approach Rb atoms are data qubits and Cs atoms are ancilla qubits. Since the two atomic species have different transition frequencies, Cs atoms could be measured without interacting with data qubits at the cost of requiring two sets of cooling lasers. In this technique, the readout beams globally address the full array allowing the parallel readout of all ancilla qubits in 15 ms. This is advantageous in that the mid-circuit readout inherently scales to a large number of ancilla qubits. The blow-away based readout used in this work limits the readout to a single mid-circuit measurement on the ancilla qubits without refilling the array; however, as the authors discuss, this technique may be modified to use non-destructive readout techniques. }

\rsub{ 
The final set of mid-circuit measurement approaches use a single atomic species and rely on data qubit shelving into dark states to prevent interaction with the readout light. This technique is known from earlier trapped ion experiemnts\cite{Riebe2004}. Our mid-circuit measurement approach is in this category as we shelve data qubits into the dark $f=3$ hyperfine manifold. After submission of our work for publication, several related preprints demonstrating mid-circuit measurements using the multi-level structure of alkaline earth like atoms appeared\cite{Lis2023,Norcia2023,Ma2023}. These manuscripts use a similar shelving-based approach to protect the data qubits from readout light. The main advantage of using alkaline earth atoms vs alkali atoms is that the data qubits do not need to be transferred out of the long-coherence computational basis which can be encoded in metastable excited states. Rather, the bright ancilla qubit states are transferred to the spin-0 ground state, which is interrogated by the readout light.}

While our demonstration did not include qubit resetting, which is required for repetitive error correction, there is a clear path towards this goal through recooling ancilla atoms in the bright state using a narrow Cs quadrupole transition (see Appendix \ref{sec.qubit_reset}). The narrow line-width of this transition combined with the large hyperfine splitting of the excited state allows cooling without the need for repumping. The coherence on the data qubits can be maintained during this cooling using dynamical decoupling techniques. Once recooled, the ancilla atoms could be reset through a combination of global microwave rotations and previously demonstrated site-selective qubit rotation techniques\cite{Graham2022}. Mid-circuit measurement and qubit resetting would also permit more efficient logical qubit initialization for error correcting codes and quantum state preparation\cite{TCLu2022}. In this application, mid-circuit measurements enable preparation of long-range entangled states faster than the limits imposed by the Lieb-Robinson bounds.   Even without resetting, this demonstration enables many semi-classical quantum computing circuits \cite{Griffiths1996,Chiaverini2005}. These circuits leverage mid-circuit measurement and feed-forward rotations to perform operations, such as a semi-classical version of a quantum Fourier transform, without using two-qubit gates. Similarly, mid-circuit measurements (without requiring qubit resetting), are an important tool for measurement-based (a.k.a. one-way) quantum computing. This computing paradigm uses mid-circuit measurements and feed-forward operations to manipulate a large, highly-entangled resource state to perform quantum computations. Neutral atom platforms are an attractive platform for measurement-based quantum computing since large qubit arrays \cite{Huft2022} and long-range controllable Rydberg interactions \cite{Saffman2010} needed for resource states have been demonstrated. Mid-circuit measurements in a neutral atom array provide a missing piece to enable measurement-based quantum computing in a neutral atom array.

This material is based upon work supported by NSF award No. 2210437, NSF Grant No. 2016136 for the Quantum Leap Challenge
Institute center Hybrid Quantum Architectures and Networks,
the U.S. Department of Energy Office of Science National Quantum
Information Science Research Centers as well as support from
the  U.S. Department of Energy, Office of Science, Office of High Energy Physics, under Grant No. DE-SC0019465.

\bibliography{atomic,saffman_refs,rydberg,qc_refs,optics}

\appendix 


\section{Experimental system}

In the following sections, we provide details of the microwave subsystem, low noise coil drivers, and the setup for laser auto relocking.

\subsection{Microwave source}
\label{sec.microwave}

\begin{figure*}[!t]
\includegraphics[width=7in]{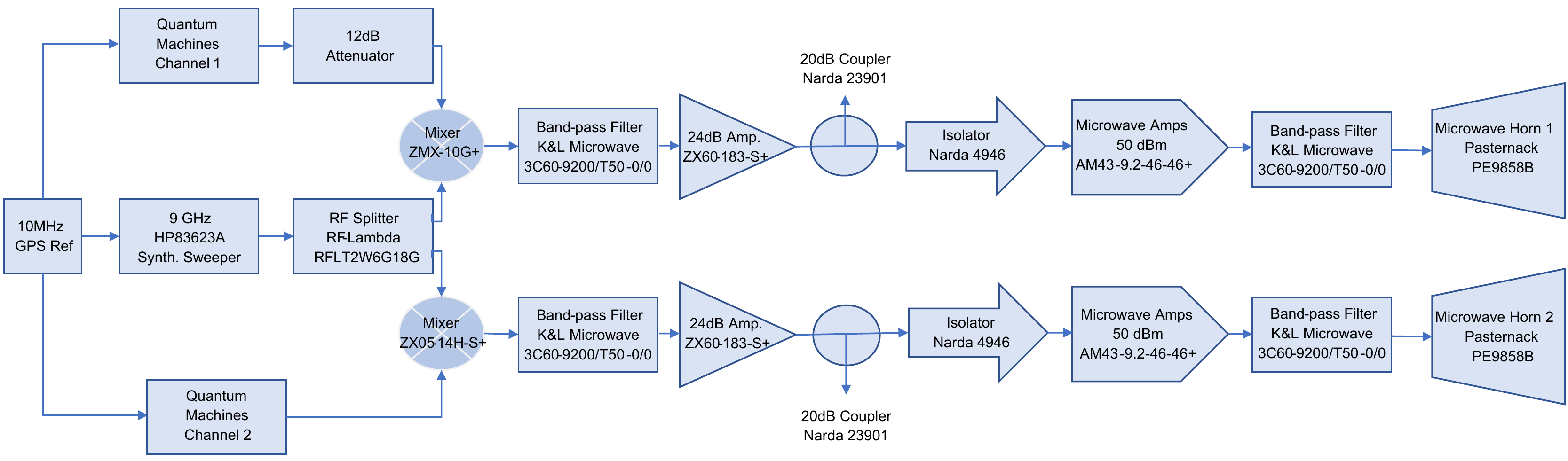}
\caption{
\label{SI_fig.uwave_layout}
Microwave signal generation diagram. Two channels from a Quantum Machines OPX at $\sim 200~\rm MHz$ are mixed with a $9~\rm GHz$ from a HP83623A. Both of these machines synchronized to a 10 MHz GPS clock. The resulting signals are filtered and amplified with a two stage amplification process. The final amplifier increases the signal power to 40 W of power before the signals are transferred to the microwave horns.
}
\end{figure*}

We used microwave rotation for all of the coherent ground state transfer operations in this experimental procedure. We designed and built a system to coherently drive two microwave horns placed in different orientations relative to the atom cloud. To generate the two signals, we mixed each of two output channels of a Quantum Machines OPX box with a 8.992631770 GHz signal provided by a Hewlett Packard HP83623A frequency synthesizer (see Fig. \ref{SI_fig.uwave_layout}). Both the OPX and HP83623A were referenced to a 10 MHz GPS clock source. After the mixer, a 9.2 GHz band-pass filter removed the carrier frequencies. The two channels were boosted with a two-stage amplification process; first pre-amplifiers increased the signals up to the 1 dB compression point of the 40 W primary amplifiers. We were able to change the relative phase between the two channels allowing limited, but precise polarization control of the microwave signal at the atoms. This control was limited by two factors. First, one of the horns could not be positioned as close to the atoms as the other horn resulting in approximately a factor of two different microwave intensities that the two horns could provide to the atoms. Second, reflections off of metal surfaces that surround the cell distorted the polarization seen by the atoms. \rsub{Note that neither of these two effects reduced the uniformity of the microwave polarization at the atom array; the microwave intensity was very isotropic due to the approximately 3 cm wavelength of the microwave light. Rather, the orientation of the microwave polarization was limited to a subset of the Bloch sphere when running the microwave horns at full power.} Despite this limitation, we were able to able to drive all microwave transitions with $>18$ kHz Rabi frequency; see Table \ref{SI_table.1} for a list of all transition Rabi frequencies. This polarization control allowed us to employ the two step qubit shelving procedure described in the main text.

\begin{table}[!t]
\caption{Observed Rabi frequencies for various hyperfine transitions. For each transition (excluding the shelving transitions), the phase between the two microwave horns was tuned to optimize the Rabi frequency. For the transitions used during the shelving pulse, $\ket{4,0}\rightarrow\ket{3,-1}$ and $\ket{3,0}\rightarrow\ket{4,-1}$, this phase difference was adjusted such that the Rabi frequency of the former transition was twice that of the latter.}\label{SI_table.1}
\centering
\begin{tabular}{|c|c| }

 \hline
   Transition  &    Rabi frequency\\
\hline
$\ket{4,4}\rightarrow\ket{3,3}$ &  99.9 kHz   \\
$\ket{3,3}\rightarrow\ket{4,3}$ &  37.9 kHz \\
$\ket{4,3}\rightarrow\ket{3,2}$ &  87.7 kHz \\
$\ket{3,2}\rightarrow\ket{4,1}$ &  18.1 kHz   \\
$\ket{4,1}\rightarrow\ket{3,0}$ &  59.9 kHz   \\
$\ket{3,0}\rightarrow\ket{4,0}$ &  62.8 kHz   \\
$\ket{3,2}\rightarrow\ket{4,2}$ &  51.8 kHz   \\
$\ket{3,1}\rightarrow\ket{4,1}$ &  56.2 kHz   \\
$\ket{4,0}\rightarrow\ket{3,-1}$ &  44.8 kHz   \\
$\ket{3,0}\rightarrow\ket{4,-1}$ &  22.4 kHz   \\
$\ket{3,-1}\rightarrow\ket{4,-1}$ &  58.4 kHz   \\

\hline
\end{tabular}
\end{table}

\subsection{Coil drivers}
\label{sec.coil_drivers}

To control the current through the coils during the experiment, we developed V-I drivers that supply current  proportionally with an input voltage. These drivers have an additional voltage input that allows for a feed-forward signal that gets added to the main target input (though this compensation was not used in this experiment). The coil drivers function with a feedback loop that controls the amount of current flowing through a pair of MOSFETs, with a buffered sense resistor supplying feedback. Large emphasis was placed into designing the system to be resilient to ground loops caused by parallel channels of the coil driver. The spectral noise density of the current of these coil drivers is less than 1 part per million above 10 Hz, and can drive up to 2 A.

\subsection{Laser auto relock}

An automatic relocking mechanism was implemented and used to maintain the lock of the 459-nm laser that is used for local single qubit shift-out operations to a high finesse stable reference cavity . The 459 nm light is obtained from a 918 nm laser provided by an M-squared Ti:Sapphire laser that is locked to an ultra-low expansion (ULE) glass cavity and is then doubled to 459 nm in a bow-tie resonator. Details of this laser system and its locking system are described in \cite{Graham2022}. The auto-relock system uses information from the spatial profile of the ULE cavity's transmission measured with a Raspberry PiCam, the total transmission power measured with a power meter, and laser frequency data measured with a Moglabs MWM wavemeter. This information is used to control the laser frequency though TCP/IP commands sent to the M Squared SolsTiS. 

\begin{figure*}[!t]
\includegraphics[width=6in]{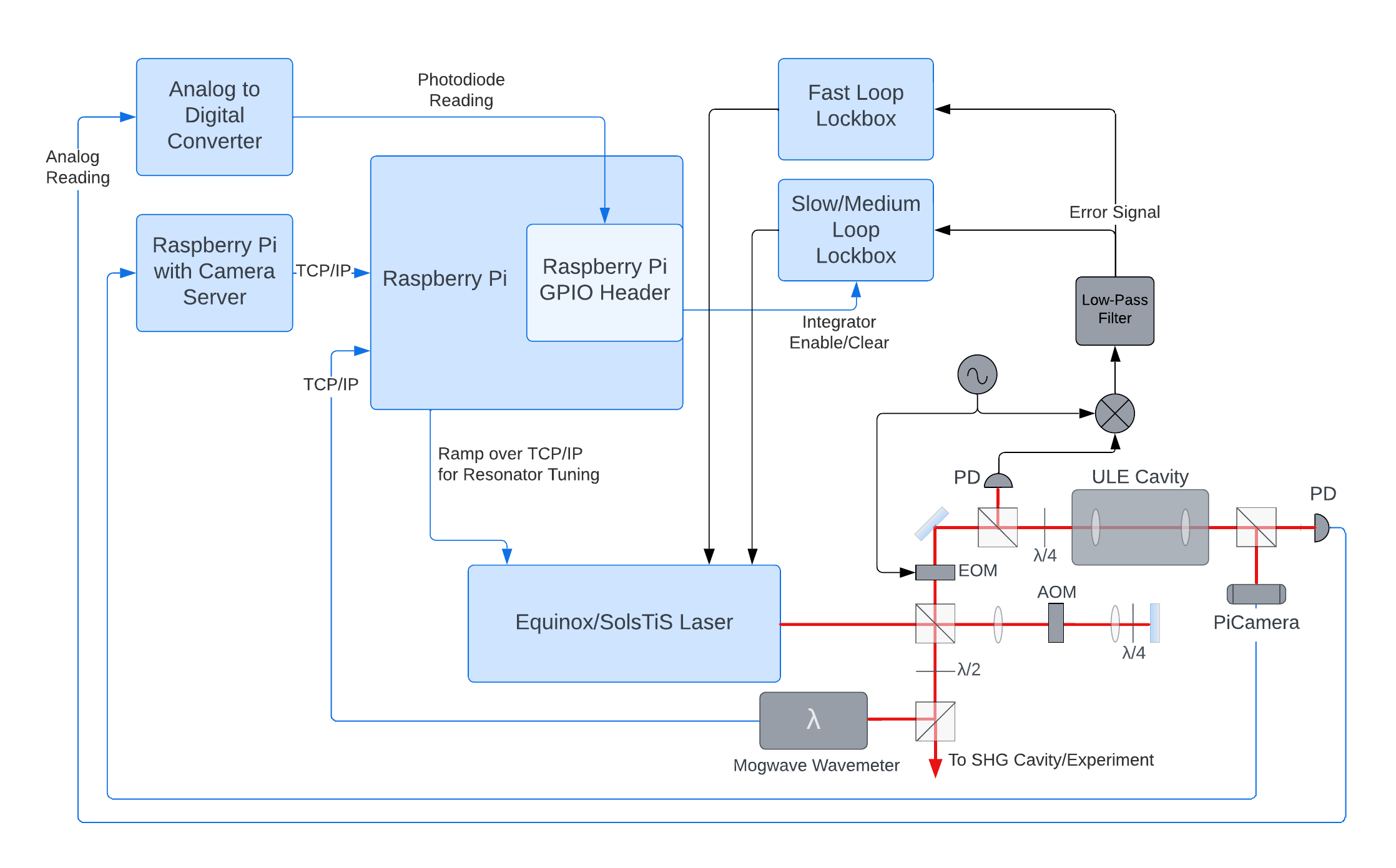}
\caption{
\label{SI_fig.autorelock}
Block diagram of the auto-relock system for the 459nm laser system described in \cite{Graham2022}. This system uses a photodiode (PD) and Raspberry-Pi Camera (PiCamera) on the transmission side of an ultra-low expansion glass cavity (ULE), as well as a Moglabs MWM wavemeter to perform automatic locking of the laser. The error signal is generated by modulating an electro-optic modulator (EOM) through the Pound-Drever-Hall technique. An acousto-optic modulator (AOM) in a double-pass configuration shifts the laser frequency to the nearest spatial mode of the ULE cavity. The 918-nm light out of the M Squared SolsTiS is converted to 459nm with a second harmonic generation (SHG) crystal/cavity (not shown). 
}
\end{figure*}

An unlock event is detectable if the transmission photodiode drops below a threshold, image analysis on the spatial mode from the camera determines that the cavity is locked to the wrong transverse mode (or not locked), or the wavemeter determines that the axial mode of the cavity has changed. The wavemeter is constantly calibrated to the ULE cavity while the laser is locked. When the laser unlocks, the frequency may drift further away from the correct transverse cavity mode by more than half of a Free spectral range (FSR). In this situation, the laser frequency is ramped in the direction of the correct transverse cavity mode. Once the laser is within half of the FSR to the previous lock point, the wavemeter does not have enough accuracy, so the laser is ramped until the correct spatial mode, determined using the camera, is found. A second higher frequency and smaller amplitude ramp is applied on top of the coarse SolsTiS ramp so that the camera shutter time integrates a wider span of the frequency allowing the spatial mode to be more easily located. Once the correct spatial mode is found, the coarse ramp is disabled and the lock integrator loops are re-enabled (the proportional and derivative loops are always enabled). This scheme has a successful relock rate of about $70\%$. It recursively attempts to lock until it succeeds or too many failed attempts have occurred (indicating that the lock parameters need to be tuned). This system does not tune error offset or loop gain, and simply re-enables loops. Eventually these parameters will saturate and the relocking system will need to be reinitialized. With this system, we  maintain  high finesse locks on the time scale of weeks and unlock events occur on time scales of days. A block diagram of this system is in Fig. \ref{SI_fig.autorelock}.

\section{Experimental pulse sequence}
\label{sec.pulse_sequence}

\begin{figure*}[!t]
\includegraphics[width=7in]{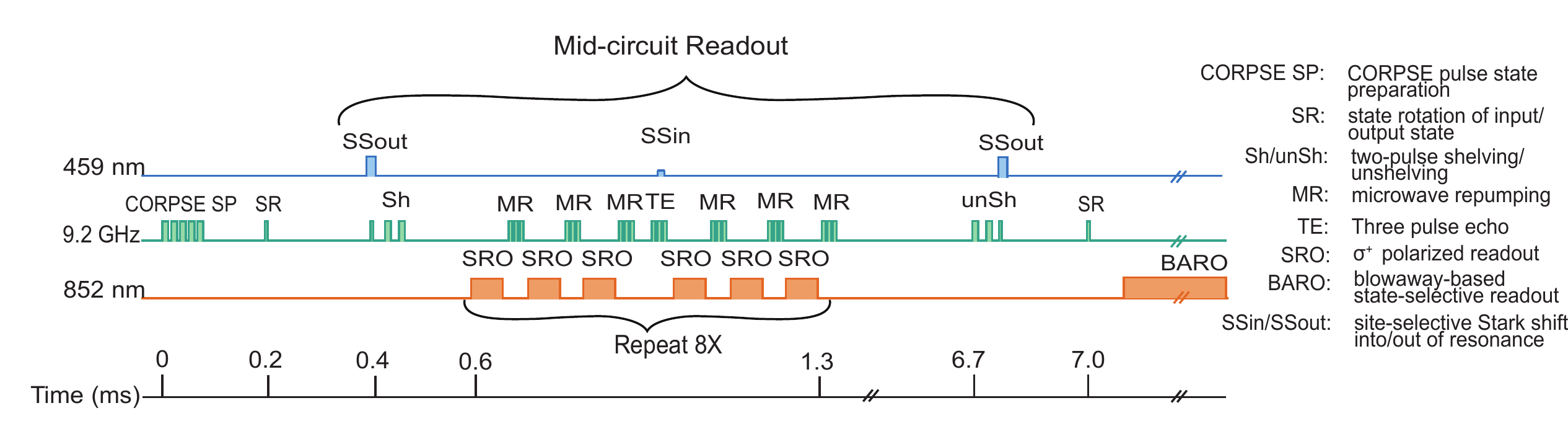}
\caption{
\label{SI_fig.circuit}
The experimental pulse sequence to characterize a mid-circuit measurement. This characterization sequence is composed of a series of pulses of microwaves to rotate qubit states, 459-nm light to provide site selective Stark shifts, and 852-nm light to cause selected atoms to fluoresce. All data and ancilla qubits begin the circuit in the $\ket{4,4}$ stretched state; five composite microwave pulses then coherently transfer the atoms to the $\ket{3,0}$ state. A microwave pulse resonant with the $\ket{3,0}\leftrightarrow\ket{4,0}$ transition prepares the input state for the mid-circuit measurement. The mid-circuit measurement begins with ramping the trap depth of the ancilla qubit up to allow more photon scattering events before the atom is lost. An extra $\pi$ pulse is applied so that any dephasing arising from the trap ramp down is echoed out when the trap is re-raised. Three pulses shelve the data qubit; during the first of these pulses, focused 459-nm light shifts the ancilla qubit out of resonance. Then, $\sigma_+$ readout light pulses are alternated with microwave pulses that repump the $\ket{3,1}$, $\ket{3,2}$, and $\ket{3,3}$ states to the $f=4$ manifold. After 3 of these cycles, a three pulse microwave echo is applied to the data qubits, then the three readout and microwave repump cycles are repeated. During the middle pulse of this echo, a low-amplitude 459-nm pulse is applied to the ancilla to allow any dark state population to be shifted into resonance. This full block is then repeated eight times in order to allow enough photons to be collected by the EMCCD camera to identify whether or not an atom was detected. The data qubits are then restored to the $\ket{3,0}$, $\ket{4,0}$ basis by running the shelving sequence in reverse order and the ancilla trap is ramped back down. The data qubits are then rotated to the output measurement basis with a microwave pulse and measured with a blowaway-based state selective readout.     
}
\end{figure*}

The experimental sequence to perform a mid-circuit measurement and characterize it is complex. This sequence can be broken into various subsections (see Fig. \ref{SI_fig.circuit}). First, five composite CORPSE pulses drive the initial, optically pumped state $\ket{4,4}$ to $\ket{3,0}$. Each CORPSE pulse is composed of a $420^{\circ}$ rotation, followed by a $300^{\circ}$ rotation with the opposite phase, followed by a $60^{\circ}$ rotation in phase with the first rotation. This sequence is insensitive to small off-resonant errors caused by magnetic field fluctuations. After the atom is prepared in $\ket{3,0}$, an additional clock state rotation could be applied to control the input state into the mid-circuit measurement. 

After completing the state preparation, we began the mid-circuit measurement process by ramping the trap powers to increase the depth of the ancilla qubit trap. This was followed by a clock-state $\pi$-pulse to both compensate the ancilla qubit for a later clock-state rotation during the shelving process and allow the small amount of data qubit dephasing during the ramp to be echoed during the ramp up at the end of mid-circuit readout process.  The data qubits were then shelved using the two step shelving process using microwave polarization control described in the main text. During the first step of the shelving process when the data qubits are shelved in $f=4$, a site-selective 459-nm laser shifts the ancilla out of the microwave resonance. During the second step of the shelving process, the data qubits shelved in $f=4$ were transferred to $f=3$ using two microwave CORPSE pulses. During this rotation, the ancilla qubit also experienced a $\pi$-rotation, which restores the qubit state prior to the mid-circuit readout start.  Once data qubits were shelved, $\sigma_+$ was applied causing any population of the ancilla qubit in $\ket{4,0}$ to fluoresce. Periodically, we applied microwave repumping pulses to prevent depumping of the ancilla bright state into the dark $f=3$ manifold; these three pulses transfer population from $\ket{3,1} \rightarrow \ket{4,1}$, $\ket{4,2} \rightarrow \ket{3,2}$, and $\ket{3,3} \rightarrow \ket{4,3}$. The data qubit coherence was preserved using periodic echoing using the following sequence of three CORPSE pulses: $\ket{3,0} \rightarrow \ket{4,0}$, $\ket{3,-1} \leftrightarrow \ket{4,0}$, and $\ket{4,0} \rightarrow \ket{3,0}$. During the duration of the mid-circuit measurement, 46 microwave repumping cycles and 8 echoes were evenly dispersed in the readout. At the end of the mid-circuit measurement, the data qubits were unshelved using the reverse of the pulse sequence that was used for shelving and the trap powers were ramped back to their initial values. A final microwave $\pi$-pulse on the clock state rotates the data qubits back to their initial state but with a extra phase between the two clock states; this phase may be compensated during the circuit compilation. After completing the mid-circuit measurement, additional clock state rotations to select the measurement basis were applied. Finally, we performed a blowaway-based state-selective measurement on the data qubits.  In all, about 246 microwave pulses were applied to perform this mid-circuit readout characterization circuit (some measurements require 1 or 2 fewer pulses), and 230 of these pulses were applied during the mid-circuit readout.  

\subsection{Alternative two-pulse shelving sequence}
\label{sec.shelving}

\begin{figure}[!t]
\includegraphics[width=3.2in]{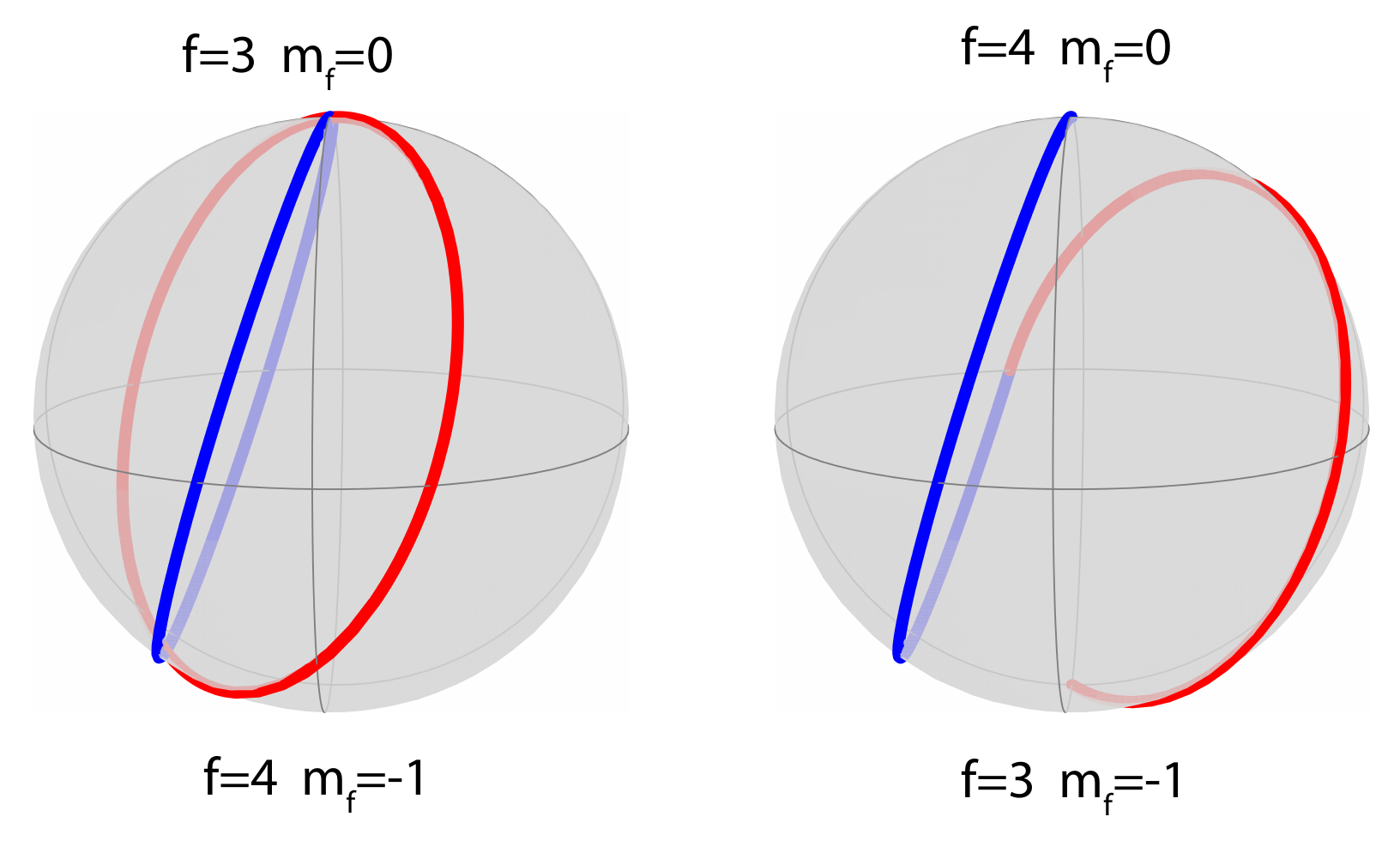}
\caption{
\label{SI_fig.shelve}
An alternative two-pulse shelving technique which does not require precise polarization control. In this sequence, differences in polarization and Clebsch-Gordan coefficients result in a difference in Rabi frequency, $\Omega_1$, of the $\ket{4,0}\rightarrow\ket{3,-1}$ transition and the Rabi frequency, $\Omega_2$, of the $\ket{3,0}\rightarrow\ket{4,-1}$ transition.  For Rabi frequency ratios of $\frac{1}{4} \leq \frac{\Omega_1}{\Omega_2}\leq \frac{3}{4}$ this difference can be exploited by tuning the pulse length and detuning to find a condition where one of the transitions (e.g. $\ket{3,0}\rightarrow\ket{4,-1}$) experience a $2\pi$ rotation while the other computational qubit state (e.g. $\ket{4,0}$) is mapped to the equator of the Bloch sphere. A second identical pulse is then applied with a relative phase with respect to the first pulse. The $\ket{3,0}$ state still undergoes a $2\pi$ rotation during the second pulse; however, with a careful selection of $\phi$, the $\ket{4,0}$ input is mapped to $\ket{3,-1}$. Pictured is an example where the microwave field has equal $\sigma_+$ and $\sigma_-$ components.
}
\end{figure}

In addition to the single pulse shelving sequence described in the main text, we designed and implemented a two-pulse shelving sequence which allows qubit shelving without precise polarization control. For this sequence we aligned the horn to have approximately even contributions of $\sigma_+$ and $\sigma_-$ polarization. We then used the difference of Clebsch-Gordan coefficients between the two transitions to design a pulse sequence to shelve the qubit. This sequence is composed of two detuned microwave pulses of the same length but different phases with respect to each other. Both pulses are designed to be detuned $2\pi$ rotations for the $\ket{3,0} \rightarrow \ket{4,-1}$ transition, so an atom in the $\ket{3,0}$ state remains in that state after both pulses (see Fig. \ref{SI_fig.shelve}). However, if an atom starts in $\ket{4,0}$, then the first microwave pulse creates an even super-position of $\ket{4,0}$ and $\ket{3,-1}$ states. By tuning the phase of the second pulse, it is possible to transfer all of the amplitude into the $\ket{3,-1}$ state. Unlike the shelving procedure described in the main text, two-pulse shelving does not require precise microwave polarization control. If the ratio of Rabi frequencies of the two transitions is between $\frac{1}{4} \leq \frac{\Omega_1}{\Omega_2}\leq \frac{3}{4}$, then there exists a combination of detuning, microwave pulse length, and phase that allows the conditions require for this shelving procedure to be met. We used this shelving sequence to obtain a shelving fidelity of $98\%$; however, the single pulse shelving in $\ket{4,0}$ and $\ket{4,-1}$ (followed by CORPSE transfer to $\ket{3,0}$ and $\ket{3,-1}$) technique allowed a higher fidelity of $99\%$ state transfer.  

\section{State preparation and measurement (SPAM) corrections}

\label{sec.state_prep}

The experiments characterizing the mid-circuit readout performance contained state preparation and measurement errors which need to be accounted for to obtain an accurate assessment of the performance. Below, we examine how such errors act on both the data qubits and ancilla qubit. We used additional experimental measurements to quantify SPAM and find corrected mid-circuit measurement performance estimates.

\subsection{SPAM correction of data qubit measurements}

An ideal mid-circuit measurement would have no effect on the data qubits and only affect the measured ancilla qubits. However, if the measurement instead acts as a known unitary operator, then this rotation can be compensated with either a corrective unitary or by accounting for this rotation in the context of the quantum circuit using a quantum circuit compiler. Errors that cannot be corrected can arise from dephasing in the shelved state or leakage outside the qubit basis due to incorrect microwave rotations and/or scattering from the readout light. We measured the likelihood of this latter error type in order to evaluate how well the mid-circuit readout protocol preserved the data qubits. Our mid-circuit measurements theoretically acted as a unitary ${\sf R}_z$ rotation on the data qubits. We measured the process fidelity of the experimental mid-circuit measurement relative to a ${\sf R}_z$ rotation to identify non-unitary errors introduced by the measurement. 

The fidelity of a single qubit quantum process with a target can be determined by inputting the states ($\ket{0}$,$\ket{1}$,$\frac{\ket{0}+\ket{1}}{\sqrt{2}}$,$\frac{\ket{0}-\ket{1}}{\sqrt{2}}$,$\frac{\ket{0}+i\ket{1}}{\sqrt{2}}$,$\frac{\ket{0}-i\ket{1}}{\sqrt{2}}$)  into the quantum process and then measuring the fidelity of the output with the corresponding target state. The average state fidelity of the six outputs is the average process fidelity. Since all target output states are pure states, we can measure the state fidelity of the experimental output with the target output by rotating the target state to $\ket{0}$ and then performing a blowaway-based state-selective readout.  For these measurements, any atoms in the correct target output state should appear bright in the readout and atoms not in the correct state should be heated out of the trap in the blowaway. However, errors in either the state preparation or measurement will affect these fidelity measurements. To compensate for these errors, we performed additional measurements to characterize them so they could be compensated for. To inform our measurements we created the error model below which details how various types of SPAM error affect the state fidelity measurements. For simplicity, we have not corrected for errors due to qubit rotations in the clock state basis during the characterization circuit. We previously characterized such errors in a similar setup to be $<3\times10^{-4}$ \cite{Graham2022}; so they did not contribute significantly to the SPAM error budget.

We first calculated the probability that an atom would be detected in the bright state during the final state-selective readout, $P(D_{B})$, based on the conditional probability of a bright output being detected given the different possible outcomes of the state prep and mid-circuit measurement. 
This probability is:
\begin{eqnarray}
\label{SI_eqn.D_B}
    P(D_{B}) & = & P(D_{B}|\ket{0}_{\rm out})P(\ket{0}_{\rm out}) \nonumber\\
                      & + & P(D_{B}|\ket{1}_{\rm out})P(\ket{1}_{\rm out}) \nonumber\\
                      & + & P(D_{B}|\ket{3,-1}_{\rm out})P(\ket{3,-1}_{\rm out}) \nonumber\\
                      & + & P(D_{B}|\ket{4,-1}_{\rm out})P(\ket{4,-1}_{\rm out}) \nonumber\\
                      & + & P(D_{B}|\epsilon_{l \rm, MCR})P(\epsilon_{l\rm, MCR}) \nonumber\\
                      & + & P(D_{B}|\epsilon_{\rm prep})P(\epsilon_{\rm prep}) \nonumber\\
                      & + & P(D_{B}|\epsilon_{l \rm,pre})P(\epsilon_{l \rm,pre}),
\end{eqnarray}
where $P(\ket{0}_{\rm out})$ is the probability that a data qubit atom was correctly initialized to $\ket{0}$ before the mid-circuit characterization circuit and was also correctly rotated to $\ket{0}$ after the circuit, $P(\ket{1}_{\rm out})$ is the probability that a data qubit atom was correctly initialized to $\ket{0}$ before the mid-circuit characterization circuit but was incorrectly rotated to $\ket{1}$ after the circuit, $P(\ket{3,-1}_{\rm out})$ is the probability that a data qubit atom was correctly initialized to $\ket{0}$ before the mid-circuit characterization circuit but was incorrectly rotated to $\ket{3,-1}$ after the circuit due to incorrect unshelving, $P(\ket{4,-1}_{\rm out})$ is the probability that a data qubit atom was correctly initialized to $\ket{0}$ before the mid-circuit characterization circuit but was incorrectly rotated to $\ket{4,-1}$ after the circuit due to incorrect unshelving, $P(\epsilon_{l\rm, MCR})$ is the probability that a data qubit atom was correctly initialized to $\ket{0}$ before the mid-circuit characterization circuit but was lost during the mid-circuit readout, $P(\epsilon_{\rm prep})$ is the probability that the atom was not properly initialized to the correct state due to either an optical pumping error or an error when rotating $\ket{4,4}$ to $\ket{0}$, and $P(\epsilon_{l \rm,pre})$ is the probability that the atom was lost from the trap before the mid-circuit characterization circuit. The conditional probabilities (e.g. $P(D_{B}|\ket{0}_{\rm out})$) represent the probability that a bright state is detected in the state selective readout given the respective outcomes of the mid-circuit measurement.  These conditional probabilities can be written in terms of how the blowaway-based state-selective readout interprets each of the outcome possibilities
\begin{eqnarray}
    P(D_{B}|\ket{0}_{\rm out}) &=& 1-\epsilon_{l \rm,post} \\
    \nonumber \\
    P(D_{B}|\ket{1}_{\rm out}) &=& \epsilon_{\rm BA}(1-\epsilon_{l \rm,post}) \approx \epsilon_{\rm BA}\\
    \nonumber \\
    P(D_{B}|\ket{3,-1}_{\rm out}) &=&  1-\epsilon_{l \rm,post} \\
    \nonumber \\
    P(D_{B}|\ket{4,-1}_{\rm out}) &=&  \epsilon_{\rm BA}(1-\epsilon_{l \rm,post}) \approx \epsilon_{\rm BA} \\
        \nonumber \\
    P(D_{B}|\epsilon_{l \rm,MCR}) & \approx &  0 \\
    \nonumber \\
    P(D_{B}|\epsilon_{\rm prep}) & = &  \epsilon_{\rm MCR, 4}(1-\epsilon_{l \rm,post}) \nonumber\\
                                 & + &\epsilon_{\rm MCR, 3}(1-\epsilon_{l \rm, post}) \nonumber\\
                                 &\approx&  \epsilon_{\rm MCR, 4}+\epsilon_{\rm MCR, 3}    \\
        \nonumber \\
    P(D_{B}|\epsilon_{l \rm,pre}) & \approx &  0, 
\end{eqnarray}

where $\epsilon_{l \rm,post}$ is the probability that an atom is lost after the mid-circuit measurement so that it is not detected during the state-selective readout, $ \epsilon_{\rm BA}$ is the probability that $\ket{1}$ is erroneously detected during the state selective measurement as a bright state due to an error in the blowaway (likely caused by the atom depumping before it could be heated out of the trap), and $\epsilon_{\rm MCR, 4}$ ($\epsilon_{\rm MCR, 3}$) is the probability that an atom improperly initialized into an $m_f$-level in $f=4$ ($f=3$) will not be blown away by the mid-circuit readout process and will occupy a state in $f=3$ going into the state-selective readout.  Note that $\epsilon_{\rm MCR, 4}$ and $\epsilon_{\rm MCR, 3}$ will both be small since the data qubits are in a shallow trap during the mid-circuit readout and the combination of readout light and microwave repumping pulses  during the mid-circuit readout will cause states with $m_f > 0$ to both heat up and collect in $f=4,m_f=4$, where the blowaway light will then heat them out of the trap (if they survived the mid-circuit readout). It is extremely unlikely for any atoms to be prepared in $m_f<0$ or $\ket{1}$ during state prep since we used stretched-state optical pumping followed by coherent transfer to $\ket{0}$. Almost all of the optical pumping error should reside in states with $\delta m_f = -1$ and $\delta m_f = -2$  away from the target state. This combined with the coherent transfer from $\ket{4,4}$ to $\ket{0}$ leaves a very small chance that the atom is erroneously prepared in $\ket{1}$ instead of $\ket{0}$. Furthermore, even if an optical pumping error resulted in the atom being erroneously prepared in $\ket{4,3}$ and it was then transferred to $\ket{4,0}$ during the CORPSE transfer segment of the state prep, the population in $\ket{4,0}$ should primarily be mapped back on to $\ket{4,0}$ at the end of the mid-circuit characterization circuit (note that since this error was originally caused by an optical pumping error, it will be incoherent  with the target input state, and therefore should not interfere and skew measurement probabilities). An atom in $\ket{4,0}$ at the end of the measurement will be heated out of the trap by the blowaway beam. The summary of this reasoning is that nearly all optical pumping and state preparation errors result in the atom being ejected from the trap before or during the blowaway of the final state-selective readout.  If the atom is lost either before or during the mid-circuit readout, then the probability that it was  detected as a bright state is negligible, since the dark histogram peak is 6.4 standard deviations from the photon count threshold needed to be classified as a bright state detection event.

The output probabilities in eq. (\ref{SI_eqn.D_B}) can be rewritten in terms of conditional probabilities of the outcomes of the state preparation process, i.e properly initialized to $\ket{0}_{\rm in}$ ($P(\ket{0}_{\rm in})$), improperly initialized to the wrong state ($\epsilon_{\rm prep}$), or lost before the mid-circuit measurement characterization circuit ($\epsilon_{l, \rm pre}$).  This probability is:

\begin{eqnarray}
    \label{SI_eq.probs}
    P(\ket{0}_{\rm out}) &=& P(\ket{0}_{\rm out}|\ket{0}_{\rm in})P(\ket{0}_{\rm in}) \nonumber \\
                         &=& P(\ket{0}_{\rm out}|\ket{0}_{\rm in})(1-\epsilon_{l \rm, pre}-\epsilon_{\rm prep}) \\
    \nonumber \\
    \label{SI-eq.p1_out}
    P(\ket{1}_{\rm out}) &=& 1- P(\ket{0}_{\rm out})- P(\ket{3,-1}_{\rm out}) \nonumber \\
                         &-& P(\ket{4,-1}_{\rm out}) - P(\epsilon_{\rm prep})-P(\epsilon_{l, \rm pre}) \nonumber \\
                         &-& P(\epsilon_{l\rm, MCR}) \nonumber \\
                         \\
    \nonumber \\
    P(\ket{3,-1}_{\rm out}) &=& P(\epsilon_{\rm sh,3}|\ket{0}_{\rm in})P(\ket{0}_{\rm in}) \nonumber \\
                         &\approx& P(\epsilon_{\rm sh,3}|\ket{0}_{\rm in}) \nonumber  \\
                         &\equiv& \epsilon_{\rm sh,3}(\psi)\\
    \nonumber \\
    P(\ket{4,-1}_{\rm out}) &=& P(\epsilon_{\rm sh,4}|\ket{0}_{\rm in})P(\ket{0}_{\rm in}) \nonumber \\
                         &\approx& P(\epsilon_{\rm sh,4}|\ket{0}_{\rm in}) \nonumber  \\
                         &\equiv& \epsilon_{\rm sh,4}(\psi)\\
    \nonumber \\    
    P(\epsilon_{\rm prep}) &\equiv& \epsilon_{\rm prep} \\
    \nonumber \\
    \label{SI_eq.loss_pre}
    P(\epsilon_{l, \rm pre}) &\equiv& \epsilon_{l, \rm pre},
\end{eqnarray}

where $\epsilon_{\rm sh,3}$ ($\epsilon_{\rm sh,4}$) represent the probability that the state was correctly prepared in $\ket{0}$ but an error in unshelving caused the state to be transferred to $\ket{3, -1}$ ($\ket{4, -1}$), $P(\ket{1}_{\rm out}|\ket{0}_{\rm in})$ is the probability that the atom was correctly prepared in $\ket{0}$ but incorrectly transferred to $\ket{1}$ due to an error in the mid-circuit calibration circuit. The conditional probability $P(\ket{0}_{\rm out}|\ket{0}_{\rm in})$ is the probability that we measure $\ket{0}$ for the output state given that the qubit was correctly initialized to $\ket{0}$; this probability represents the SPAM corrected state fidelity that we want to calculate.

We can further simplify eq. (\ref{SI-eq.p1_out}) by plugging in the other probabilities from eqs. (\ref{SI_eq.probs})-(\ref{SI_eq.loss_pre}):
\begin{eqnarray}
        P(\ket{1}_{\rm out}) &=& 1- P(\ket{0}_{\rm out}|\ket{0}_{\rm in})(1-\epsilon_{l \rm, pre}-\epsilon_{\rm prep})\nonumber \\
        &-& \epsilon_{\rm sh,3}(\psi) - \epsilon_{\rm sh,4}(\psi) -  \epsilon_{\rm prep}-\epsilon_{l, \rm pre} \nonumber \\
        &-&P(\epsilon_{l\rm, MCR})
\end{eqnarray}

Plugging these probabilities and conditional probabilities into eq. (\ref{SI_eqn.D_B}) yields
\begin{eqnarray}
    P(D_{B}) & \approx &  (1-\epsilon_{l \rm,post})P(\ket{0}_{\rm out}|\ket{0}_{\rm in}) \nonumber\\
                    & \times &(1-\epsilon_{l \rm, pre}-\epsilon_{\rm prep}) \nonumber\\
                      & + & \epsilon_{\rm BA}P(\ket{1}_{\rm out})  \nonumber\\
                      & + &  (1-\epsilon_{l \rm,post}) \epsilon_{\rm sh,3}(\psi) \nonumber\\
                      & + & \epsilon_{\rm BA}\epsilon_{\rm sh,4}(\psi) \nonumber\\
                      & + &  (\epsilon_{\rm MCR, 4}+\epsilon_{\rm MCR, 3})\epsilon_{\rm prep}.
\end{eqnarray}

All $\epsilon$ terms represent small error probabilities, so approximating to only first order errors, we can further simplify to

\begin{eqnarray}
\label{SI_eqn.DB_simp}
    P(D_{B}) & \approx &  P(\ket{0}_{\rm out}|\ket{0}_{\rm in})\nonumber\\
                      & \times &(1-\epsilon_{l \rm, pre}-\epsilon_{l \rm,post}-\epsilon_{\rm prep}) \nonumber\\
                      & + & \epsilon_{\rm BA}\left[ 1- P(\ket{0}_{\rm out}|\ket{0}_{\rm in}) \right]+\epsilon_{\rm sh,3}(\psi) .
\end{eqnarray}

Then solving for the SPAM corrected error, $ P(\ket{0}_{\rm out}|\ket{0}_{\rm in})$, yields

\begin{equation}
\label{SI_eqn.SPAM_C}
   P(\ket{0}_{\rm out}|\ket{0}_{\rm in})  \approx  \frac{P(D_{B})-\epsilon_{\rm BA}- \epsilon_{\rm sh,3}(\psi)}{1-\epsilon_{l \rm, pre}-\epsilon_{l \rm,post}-\epsilon_{\rm prep}-\epsilon_{\rm BA}}.
\end{equation}

These values can be determined from three different measurements. Two measurements are needed to determine the state preparation error probability, $\epsilon_{\rm prep}$. For convenience, we break state preparation errors into two parts, the probability that a state is incorrectly prepared in $\ket{f=3, m_f \neq 0}$, $\epsilon_{3, \rm prep}$, and the probability that a state is incorrectly prepared in $\ket{f=4, m_f \neq 0}$, $\epsilon_{4, \rm prep}$. Note, that for this analysis, we assume, for reasons listed above, that the probability of erroneously preparing $\ket{4,0}$ is negligible.  To determine, $\epsilon_{3, \rm prep}$, we prepare $\ket{3,0}$ by optically pumping into $\ket{4,4}$, then coherently transfer the atoms to $\ket{3,0}$ using CORPSE pulses. We then perform a state-selective measurement by performing a blowaway and measure the array occupation.  The probability that a data qubit is retained, $ R_{3\rm prep}$, is then

\begin{eqnarray}
     R_{3\rm prep}&=&(1-\epsilon_{l \rm pre}-\epsilon_{l \rm post}) \nonumber \\
                     &-&\left[1-\epsilon_{\rm BA}(1-\epsilon_{l \rm pre}-\epsilon_{l \rm post}) \right] \epsilon_{4, \rm prep} \nonumber \\
                     &\approx& 1-\epsilon_{l \rm pre}-\epsilon_{l \rm post}-\epsilon_{4, \rm prep}.
\end{eqnarray}

If we perform the same experiment with the addition of a (approximately perfect) microwave rotation from $\ket{3,0}$ to $\ket{4,0}$ then we find a retention probability, $ R_{4\rm prep}$ of: 

\begin{eqnarray}
     R_{4\rm prep}&=&(1-\epsilon_{l \rm pre}-\epsilon_{l \rm post})\epsilon_{3, \rm prep} \nonumber \\
                     &-&\epsilon_{\rm BA}(1-\epsilon_{l \rm pre}-\epsilon_{l \rm post})(1-\epsilon_{3, \rm prep}) \nonumber \\
                     &\approx& \epsilon_{3, \rm prep} + \epsilon_{\rm BA}.
\end{eqnarray}
We observe that the denominator in eq. (\ref{SI_eqn.SPAM_C}) can be written as $R_{3\rm prep}-R_{4\rm prep}$. 

One final measurement is needed to determine $\epsilon_{\rm sh,3}(\psi)$. We do not have a straightforward way of directly determining this value since this error  cannot be easily separated from the other errors arising from the mid-circuit measurement. Furthermore, this error is state dependent, so we would need to disentangle it for every input. What we can do is find an average upper-bound on this error. This will make our SPAM-corrected fidelity a conservative estimate since larger $\epsilon_{\rm sh,3_{\rm ave}}$ results in a lower average fidelity.  We can take this estimate from the minimum value on the Ramsey curve  shown in Fig. \ref{fig.Data}a. To estimate how this data relates to $\epsilon_{\rm sh,3}(\psi=x)$ (where $\ket{x}=(\ket{0}+\ket{1})/\sqrt{2}$), we can use eq. (\ref{SI_eqn.DB_simp}) and set $P(\ket{0}_{\rm out}|\ket{0}_{\rm in})=0$. Then solving for $\epsilon_{\rm sh,3}(\ket{\psi}=\ket{x})$, we find

\begin{equation}
      \epsilon_{\rm sh,3_{\rm ave}}=\epsilon_{sh,3}(\ket{\psi}=\ket{x}) \leq P(D_{B,\rm min})-\epsilon_{\rm BA},
\end{equation}

where $P(D_{B,\rm min})$ is the minimum retention probability of the Ramsey curve. Note that this is a conservative over-estimation of the $\epsilon_{sh,3}(\ket{\psi}=\ket{x})$ because various errors in the mid-circuit readout such as dephasing might increase $P(\ket{1}_{\rm out}|\ket{0}_{\rm in})$. An over estimation of $\epsilon_{\rm sh,3}(\ket{\psi}=\ket{x})$ is conservative for the fidelity measurement because this is the probability that an error during the mid-circuit measurement might make the fidelity look higher than it actually is, and is subtracted off the true fidelity to find the SPAM corrected fidelity. Note that this value should also be equivalent to the $\epsilon_{\rm sh,3 \rm ave}$, it gives equal weighting to $\ket{0}$ and $\ket{1}$ inputs and in the average process fidelity measurement, the two clock state inputs have equal weighting. Note that this type of error is caused by unshelving errors causing leakage out of the qubit basis and will not depend on a relative phase between clock state inputs.

We can now rewrite eq. (\ref{SI_eqn.SPAM_C}) in terms of experimentally measured quantities:

\begin{equation}
       P(\ket{0}_{\rm out}|\ket{0}_{\rm in}) \geq \frac{P(D_B)-P(D_{B,\rm min})}{R_{3,\rm prep}-R_{4,\rm prep}}.
\end{equation}
We have  measured the input values $P(D_{B,\rm min})=0.01(1)$, $R_{3,\rm prep}=0.970(3)$, and $R_{4,\rm prep}=0.014(2)$. A summary of all data qubit raw and SPAM corrected fidelities is given  in Table \ref{SI_table.2}.  Averaging this quantity for all 6 inputs yields the average process fidelity for the data qubits in the mid-circuit measurement. 

\begin{table}[!t]
\caption{Raw and state preparation and measurement (SPAM) corrected fidelity measurements of the data qubits. We have assumed that $\epsilon_{\rm sh,3}(\ket{\psi})$, the probability that the mid-circuit measurement incorrectly moves an input state $\ket{\psi}$ correctly prepared in the computational qubit basis to a state in $f=3$ where $m_f \neq 0$, is the same for the two inputs $\ket{0}$ and $\ket{1}$.  Note that this assumption is not needed for the average fidelity reported in the final row since on average the inputs have an equal weighting of both terms.}\label{SI_table.2}
\centering
\begin{tabular}{|c|c|c| }

 \hline
   Input state  &   Raw fidelity &   SPAM-corrected fidelity\\
\hline
$(\ket{0}+\ket{1})/\sqrt{2}$ &  93.0(8)\% &  96.2(1.2)\%\\
$(\ket{0}-\ket{1})/\sqrt{2}$ &  94.1(8)\% & 97.4(1.3)\%\\
$(\ket{0}+i\ket{1})/\sqrt{2}$ &  93.4(9)\% & 96.6(1.3)\%\\
$(\ket{0}-i\ket{1})/\sqrt{2}$ &  93.4(1.0)\% & 96.6(1.4)\%\\
$\ket{0}$ &  94.0(4)\% & 97.2(1.3)\%\\
$\ket{1}$ &  94.6(3)\%  & 97.9(1.3)\%\\
\hline
Average & 93.8(3)\% & 97.0(5)\%\\

\hline
\end{tabular}
\end{table}

\subsection{Correction of state preparation errors on  ancilla qubits}

A similar process may be used to compensate mid-circuit measurements on the ancilla qubit for state preparation errors. The raw ancilla qubit measurement characterization was performed in two experiments.  In the first, $\ket{0}$ was prepared, and then a mid-circuit measurement was performed. In the ideal case, $\ket{0}$ would be detected as a dark state by the mid-circuit measurement procedure. We can write the probability of a dark state detection as:
\begin{eqnarray}
 \label{SI_eqn.P1D}
    P_1(D) & = & P(D|\ket{0})P_1(\ket{0}_{\rm in}) \nonumber\\
                   & + & P(D|\epsilon_{l \rm, pre})P(\epsilon_{l \rm, pre}) \nonumber\\
                   & + & P(D|\epsilon_{\rm prep})P(\epsilon_{\rm prep}),
\end{eqnarray}
where $P_1(D)$ is the probability that a dark state is registered by the mid-circuit measurement in this first experiment, $P(D|\ket{0})$ is the probability that a dark state is detected given a $\ket{0}$ input, $P(\ket{0}_{\rm in})$ is the probability that the state $\ket{0}$ is correctly prepared in this first experiment, $P(D|\epsilon_{l \rm, pre})$ is the probability of registering a dark state given loss of the atom before the mid-circuit measurement, $P(\epsilon_{l \rm, pre})$ is the probability that the atom is lost before the mid-circuit measurement, $P(D|\epsilon_{\rm prep})$ is the probability of registering a dark state if the atom in not prepared in the correct state, and $P(\epsilon_{\rm prep})$ is the probability of a state preparation error. Note that for this analysis, for the same reasons as listed in the previous section, we have assumed that a state preparation error resulting in the $\ket{1}$ state  has negligible probability. 

A second experiment was then performed by preparing the input $\ket{0}$, then performing a clock state rotation to $\ket{1}$, and then performing the mid-circuit measurement. Ideally, a $\ket{1}$ input would be registered as a bright state by the mid-circuit readout procedure. We can write the probability of detecting a bright state in this case as:
\begin{eqnarray}
    \label{SI_eqn.P2B}
    P_2(B) & = & P(B|\ket{1})P_2(\ket{1}_{\rm in}) \nonumber\\
                   & + & P(B|\epsilon_{\rm{loss}})P(\epsilon_{\rm{loss}}) \nonumber\\
                   & + & P(B|\epsilon_{\rm prep})P(\epsilon_{\rm prep}),
\end{eqnarray}
where $P_2(B)$ is the probability that a bright state is registered by the mid-circuit measurement in this second experiment, $P(B|\ket{1})$ is the probability that a bright state is detected given a $\ket{1}$ input, $P(\ket{1}_{\rm in})$ is the probability that the state $\ket{1}$ is correctly prepared in this second experiment, $P(B|\epsilon_{\rm{loss}})$ is the probability that a bright state is registered even though the atom is lost prior to the mid-circuit measurement, and $P(B|\epsilon_{\rm prep})$ is the probability that a dark count is registered when there is a state preparation error. Note that since we assumed that the probability of erroneously preparing $\ket{1}$ in the first experiment was negligible and we have assumed   nearly perfect clock state rotations. 
It follows that the probability of erroneously preparing $\ket{0}$ in the second experiment is negligible. For the same reasons we may identify that
\begin{equation}
P_1(\ket{0}_{\rm in})=P_2(\ket{1}_{\rm in}) = 1-\epsilon_{l,\rm pre}-\epsilon_{\rm prep}.    
\end{equation}
We may also identify that:
\begin{equation}
    P(D|\epsilon_{l \rm, pre}) = 1-P(B|\epsilon_{l \rm, pre})\equiv 1-\epsilon_{l,\rm xtalk},
\end{equation}
where $\epsilon_{l,\rm xtalk}$ represents the probability the a lost atom will be registered as a bright state. Note that $\epsilon_{l,\rm xtalk}$ should be small since an atom that is lost from the trap will not fluoresce and such mislabeling arises only from overlap in the histogram distributions for no atom and an atom detected in the bright state.  Furthermore, we observe that
\begin{equation}
    P(B|\epsilon_{\rm prep}) = 1-P(D|\epsilon_{\rm prep}) \equiv 1-\epsilon_{D,\rm prep},
\end{equation}
where $\epsilon_{D,\rm prep}$ represents the probability that an atom prepared in the wrong state will be registered as a dark state. This probability should also be small because atoms prepared in $\ket{4,m_f>0}$ will be pumped to $\ket{4,4}$ by the readout light just as atoms in $\ket{1}$ are. Similarly, the periodic microwave repumping will result in atoms incorrectly prepared in $\ket{3,m_f>0}$ undergoing transitions to the $f=4$ manifold, where they too will be pumped to $\ket{4,4}$ by the readout light.  We may now simplify eq. (\ref{SI_eqn.P1D}) to
\begin{eqnarray}
 \label{SI_eqn.P1D_simp}
    P_1(D) & = & P(D|\ket{0})P_1(\ket{0}_{\rm in}) \nonumber\\
                   & + & (1-\epsilon_{l,\rm xtalk})\epsilon_{l \rm, pre} \nonumber\\
                   & + & \epsilon_{D,\rm prep}\epsilon_{\rm prep} \nonumber\\
                   & \approx & P(D|\ket{0})(1-\epsilon_{l \rm, pre}-\epsilon_{\rm prep}) \nonumber \\
                   & + & \epsilon_{l \rm, pre}
\end{eqnarray}
and eq. (\ref{SI_eqn.P2B}) to
\begin{eqnarray}
    \label{SI_eqn.P2B_simp}
    P_2(B) & = & P(B|\ket{1})P_2(\ket{1}_{\rm in}) \nonumber\\
                   & + & \epsilon_{l,\rm xtalk}\epsilon_{\rm{l \rm, pre}} \nonumber\\
                   & + & (1-\epsilon_{D,\rm prep})\epsilon_{\rm prep} \nonumber\\
                   & \approx & P(B|\ket{1})(1-\epsilon_{l \rm, pre}-\epsilon_{\rm prep}) \nonumber\\
                   & + & \epsilon_{\rm prep}.
\end{eqnarray}
In these equations, we have kept first order terms in $\epsilon$. We can now solve for the state preparation corrected fidelities $P(D|\ket{0})$ and $P(B|\ket{1})$:
\begin{eqnarray}
    \label{SI_eqn.P2B_simp2}
    P(D|\ket{0}) & = &\frac{P_1(D)-\epsilon_{l \rm, pre}}{1-\epsilon_{l \rm, pre}-\epsilon_{\rm prep}} \\
                   \nonumber\\
    P(B|\ket{1}) & = &\frac{P_2(B)-\epsilon_{\rm prep}}{1-\epsilon_{l \rm, pre}-\epsilon_{\rm prep}} \\              
\end{eqnarray}

We wish to express $P(D|\ket{0})$ and $P(B|\ket{1})$, the compensated detection probabilities, in terms of experimentally measured quantities. To do this we performed measurements to estimate $P(\epsilon_{l \rm, pre})$, $P(\epsilon_{\rm prep})$.  These were characterized in four experiments. Two of the experiments, $ R_{3\rm prep}$ and $R_{4\rm prep}$, were the same as those performed in the previous section. In addition, we performed an experiment to find the loss present before the mid-circuit measurement by performing a measurement on the atoms followed by a second measurement. The probability that the atom is detected in the second measurement is
\begin{equation}
     R_{\rm base}=1-\epsilon_{l \rm pre}-\epsilon_{l \rm post}.
\end{equation}
If we assume that $\epsilon_{l \rm pre} \approx \epsilon_{l \rm post}$ then:
\begin{equation}
     \epsilon_{l, \rm pre} \approx \frac{1-R_{\rm base}}{2}
\end{equation}

To disentangle blowaway from $R_{3\rm prep}$ and $R_{4\rm prep}$, we performed a measurement of the blowaway by preparing $\ket{4,4}$ with optical pumping and then performing a blowaway and readout. During our optical pumping we had an excess of repump light. We also measured a depumping to pumping time constant ratio  $>200$. These 
two facts together imply that there will be a negligible amount of atom population left in the $f=3$ manifold. The resulting atom retention probability for this experiment is
\begin{equation}
     R_{\rm BA} = \epsilon_{\rm BA}(1-\epsilon_{l, \rm pre}-\epsilon_{l, \rm post})\approx \epsilon_{\rm BA}.
\end{equation}
We used all four experimental results to determine $\epsilon_{\rm prep}$ as
\begin{equation}
    \epsilon_{\rm prep} = R_{4 \rm prep}-R_{3 \rm prep}+R_{\rm base}-R_{\rm BA}
\end{equation}
We can now express the state preparation corrected mid-circuit measurement fidelities as
\begin{eqnarray}
    P(D|\ket{0}) & = & \frac{P_1(D)-\frac{1}{2}(1-R_{\rm base})}{\frac{1}{2}-R_{4,\rm prep}+R_{3,\rm prep}-\frac{R_{\rm base}}{2} + R_{\rm BA}} \nonumber\\
    \\
    P(B|\ket{1}) & = & \frac{P_2(B)-R_{4, \rm prep}+R_{3, \rm prep}-R_{\rm base}+R_{\rm BA}}{\frac{1}{2}-R_{4,\rm prep}+R_{3,\rm prep}-\frac{R_{\rm base}}{2} + R_{\rm BA}}. \nonumber\\
\end{eqnarray}
Plugging in measured values ($P_1(D)=0.936(5)$, $P_2(B)=0.943(5)$, $R_{\rm base}=0.977(5)$, $R_{4,\rm prep}=0.020(2)$, $R_{3,\rm prep}=0.977(5)$, $R_{\rm BA}=0.005(2)$), we find state preparation compensated detection fidelities of $P(D|\ket{0})=0.949(8)$ and $P(B|\ket{1})=0.953(11)$.

\section{\rsub{Reaching fault tolerance thresholds}}
\label{sec.fault_tolerance}

\rsub{ 
The mid-circuit measurement fidelities presented here are below the requirements for fault-tolerant quantum circuit operations.  However, the presented results  do not represent fundamental performance limitations but rather the impact of technical imperfections and the capabilities of the current experimental apparatus. In this section, we discuss the experimental improvements needed to achieve fault tolerance and discuss the ultimate limitations of this measurement technique.  We have divided the analysis  into three sections: 1. Magnetic noise, 2. Zeeman state rotations, and 3. Ancilla Readout.}

\subsection{\rsub{Magnetic and vector Stark shift noise}}
\label{subsec.magnetic_noise}
\rsub{ 
The chief limiting factor in this demonstration is magnetic noise from stray fields in the lab and residual vector Stark shifts from imperfect trap polarization. These two noise sources cause broadening of magnetically sensitive transitions. This broadening reduces the fidelity of transitions between magnetically sensitive Zeeman states and also reduces the coherence of information stored in the magnetically sensitive shelved states. In this manuscript, we have partially compensated for these imperfections by using composite CORPSE pulses to reduce the frequency sensitivity of these transitions. We have also used periodic decoupling pulses on the data qubits when they are shelved in magnetically sensitive states during the mid-circuit readout.  These strategies were required to achieve the fidelity presented in the main text of the manuscript but were not able to fully compensate for the broadening due to limitations in the microwave Rabi frequency. If magnetic noise and vector Stark shifts in the trap were fully compensated, then the fidelity of transitions between magnetically sensitive states will approach that of the computational qubit states ($\ket{f=3, m_f=0}$ and $\ket{f=4, m_f=0}$), which have been measured in previous iterations of this experiment to have an average global Clifford fidelity of 99.98\% with site-selective Clifford rotations of 99.3\% fidelity \cite{Graham2022}.  Similarly elimination of magnetic and vector Stark noise would extend the $T_2$ coherence time to the computational qubit coherence time.  The coherence time of a qubit in the computational basis state was measured to be approximately 1 s in a previous iteration of this experiment \cite{Graham2022}). }

\rsub{
Significant reduction of the magnetic field noise can be achieved with various experimental upgrades such as adding $\mu$-metal shielding, and using permanent magnets for the bias fields applied during the circuit \footnote{Electro-magnets could be used during other phases of the experimental sequence  such as atom loading and cooling}. Ruster et al., have previously used such techniques to reduce the magnetic field to $\leq 2.7 \times 10^{-12}$ T \cite{Ruster2016}; roughly two orders of magnitude lower than our estimated magnetic field noise. The vector Stark shift from the trap light can  be largely reduced by redesigning the optical path so that a high-extinction-ratio polarizer is added to the optical path  before  focusing onto the atoms. Some residual circular polarization will still be present due to vector diffraction effects \cite{Saffman2005a,Thompson2013}. Noise effects of this residual polarization can be reduced through improved cooling and trap power stabilization.  With such improvements, we believe that it is realistic to obtain an echoed $T_2$ time of at least one second for the stored data qubits.}

\subsection{\rsub{ Zeeman state rotations}}
\label{subsec.qubit_rotations}

\rsub{ 
Accurate control and transfer of qubit amplitudes between Zeeman substates is needed for shelving and un-shelving of  data qubits and measurement of ancilla qubits. 
As discussed in App. \ref{subsec.magnetic_noise}, the fidelity of these operations can be significantly improved by reducing magnetic field and vector Stark shift noise. 
Gates on data qubits can also be improved by increasing the Rabi frequency of the rotations, better stabilization of the microwave field, and with composite pulse sequences. To increase the Rabi frequency, the microwave intensity on the atoms needs to be increased. This can be achieved using phased array antennas, higher power amplifiers, or mounting a microwave antenna inside the cell. Microwave driven neutral atom gates with infidelity below $10^{-4}$ in a large array were recently demonstrated\cite{Nikolov2023}.}   

\rsub{ 
In order to preserve the state of ancilla qubits while the data qubits are shelved we have employed a local differential Stark shift from a 459 nm beam acting as a ``shift-out" operation to prevent ancillas from being affected by the shelving microwave pulses. 
this leads to two sources of error on the ancilla state. Photon scattering which changes the state, and incomplete shift-out leading to a partial rotation of the state. Note that changes to the phase of the ancilla state are not problematic since the phase does not affect measurement probabilities in the computational basis. }

\rsub{ 
The small ancilla scattering probability during the shift-out pulse, averaged over the computational basis states, can be expressed as  
\begin{equation}
    p_{\rm scat}\simeq\frac{1}{2}\gamma_{7\rm p} t
    \left(  \frac{|\Omega_{459}|^2}{4\Delta_{7\rm p}^2}+\frac{|\Omega_{459}|^2}{4(\Delta_{7\rm p}-\omega_{\rm q})^2}
\right),
\end{equation}
where $\gamma_{7p}$ is the linewidth of the $7p_{1/2}$ state, $t$ is the pulse duration, $\Omega_{459}$ is the Rabi frequency for the $6s_{1/2}\ket{f,0} \rightarrow 7p_{1/2}\ket{f',0}\, (f\ne f')$ transition, 
and $\Delta_{7p}$ is the detuning of the 459-nm beam from the $f=4 \rightarrow 7p_{1/2}$ center of mass. This expression assumes the pulse is long compared to the $7p_{1/2}$ lifetime, so $t\gamma_{\rm 7p}\gg 1$. Since the detuning of -24 GHz is large compared to the 377 MHz $7p_{1/2}, f=3 - 4$ hyperfine splitting we ignore small corrections from the hyperfine shifts.  In the limit of large detuning of the 459 nm light $(\Delta_{\rm 7p}\gg \omega_q) $ the scattering error is 
\begin{equation}
p_{\rm scat}\simeq \gamma_{\rm 7p} t \frac{|\Omega_{459}|^2}{4\Delta_{\rm 7p}^2} 
=\frac{\pi\gamma_{\rm 7p}}{4|\Omega_\mu|}  \frac{|\Omega_{459}|^2}{\Delta_{\rm 7p}^2} 
\label{eq.scaterror}
\end{equation}
for a shift-out pulse of duration corresponding to a $\pi$ microwave pulse with Rabi frequency $\Omega_\mu.$ }

\rsub{ 
The shift-out light imparts  light shifts to the computational basis states, and the states involved in the first step of the shelving process (see Fig. \ref{fig.1}b), of  
\begin{eqnarray}
     \Delta_{4,0}= \Delta_{4,-1}&=&      \frac{|\Omega_{459}|^2}{4\Delta_{7p}},\nonumber\\
        \Delta_{3,0}= \Delta_{3,-1}&=&\frac{|\Omega_{459}|^2}{4(\Delta_{7p}-\omega_{\rm q})}.\nonumber
\end{eqnarray}
The first shelving pulse drives a $2\pi$ rotation of the unshifted data qubits on $\ket{4,0}\leftrightarrow \ket{3,-1}$ at Rabi frequency $\Omega_{\mu}/2$ and a $\pi$ rotation on $\ket{3,0}\leftrightarrow \ket{4,-1}$ at Rabi frequency $\Omega_\mu$. An ancilla qubit that is in the state $\ket{\psi}=c_0\ket{4,0}+c_1\ket{3,0}$ sees the same microwave driven rotations, but with a detuning of $\Delta_{\rm DLS}=\Delta_{4,0}-\Delta_{3,0}.$
In the large detuning limit the differential light shift is $\Delta_{\rm DLS}=\omega_q|\Omega_{459}|^2/(4\Delta_{\rm 7p}^2)$. We can therefore express the scattering error as 
\begin{equation}
p_{\rm scat}\simeq \pi\frac{\gamma_{\rm 7p}\Delta_{\rm DLS}}{\omega_{\rm q}|\Omega_\mu| }. 
\label{eq.scaterror2}
\end{equation}}

\rsub{ 
Since the first shelving pulse does not directly couple the computational basis states we do not need to consider any interference effects when accounting for the effect of the shifted-out pulse. The change of the populations due to the detuned pulses is 
\begin{widetext}
\begin{eqnarray}
    |c_0|^2&\rightarrow& \left|\cos\left(\pi \frac{\sqrt{|\Omega_\mu|^2+4\Delta_{\rm DLS}^2}}{|\Omega_\mu|}  \right)-i \frac{2\Delta_{\rm DLS}}{\sqrt{|\Omega_\mu|^2+4\Delta_{\rm DLS}^2} }\sin\left(\pi \frac{\sqrt{|\Omega_\mu|^2+4\Delta_{\rm DLS}^2}}{|\Omega_\mu|}  \right) \right|^2|c_0|^2, \nonumber\\
     |c_1|^2&\rightarrow &\left|\cos\left(\pi \frac{\sqrt{|\Omega_\mu|^2+\Delta_{\rm DLS}^2}}{2|\Omega_\mu|}  \right)-i \frac{\Delta_{\rm DLS}}{\sqrt{|\Omega_\mu|^2+\Delta_{\rm DLS}^2} }\sin\left(\pi \frac{\sqrt{|\Omega_\mu|^2+\Delta_{\rm DLS}^2}}{2|\Omega_\mu|}  \right) \right|^2|c_1|^2 \nonumber.
\end{eqnarray}
\end{widetext} 
After expanding in the small parameter $\epsilon=|\Omega_\mu/\Delta_{\rm DLS}|$ we find the population errors at leading order are 
\begin{subequations}  
\begin{eqnarray}
    |c_0|^2&\rightarrow& \left[ 1-\frac{\epsilon^2}{8}+ \frac{\epsilon^2}{8}\cos(4\pi/\epsilon+\pi\epsilon/2)\right]|c_0|^2, \\
     |c_1|^2&\rightarrow &\left[1-\frac{\epsilon^2}{2}+ \frac{\epsilon^2}{2}\cos(\pi/\epsilon-\pi\epsilon/2) \right]|c_1|^2 .
\end{eqnarray}
\label{eq.c0c1}
\end{subequations}
 Averaging over the basis states and the error oscillations we take the rotation error to be $p_{\rm rot}=\frac{5}{16}\epsilon^2.$ The total error probability for the ancilla qubits is then
\begin{equation}
    p=\frac{1}{3}p_{\rm scat}+p_{\rm rot}\simeq \frac{\pi}{3}\frac{\gamma_{\rm 7p}}{\omega_{\rm q}\epsilon }+\frac{5}{16}\epsilon^2.
\end{equation}
The additional factor of $1/3$ in the contribution from the scattering error follows from accounting for the branching ratios of the scattering process. Following the full sequence of microwave pulses, and pumping of the ancilla $f=4$ states to the stretched state for readout, we find that for an ancilla initially in $\ket{4,0}$ (the ancilla dark state) errors result from photon scattering that changes the state to $\ket{4,1}, \ket{3,\pm1}$. These events have statistical weight of $11/24$. For an ancilla initially in $\ket{3,0}$ (the ancilla bright state) errors result from photon scattering that changes the state to $ \ket{4,-1}$. These events have statistical weight of $5/24$. Averaging over the ancilla states we get a factor of $1/3.$ We note that these branching ratios apply to the situation where the $7p_{1/2}$ detuning is small compared to the $7p$ fine structure splitting. In the opposite limit Raman events are strongly suppressed\cite{Cline1994} which would be even more favorable. However the required optical power would be prohibitive. }

\rsub{ 
Minimizing the error probability with respect to $\epsilon$ we find that the minimum achievable error is obtained for 
$$ 
\epsilon=\left(\frac{2\pi}{15}\frac{\gamma_{7p}}{\omega_{\rm q}}\right)^{1/3}
$$
which leads to a minimum error probability of 
$$
p_{\rm min}=\left(\frac{15\pi^2 }{64}\frac{\gamma_{7p}^2}{\omega_{\rm q}^2}\right)^{1/3}.
$$
Using $\gamma_{\rm 7p }=1/(165 ~\rm ns)$\cite{Toh2019}  we find $p_{\rm min}=0.29\%$. 
}

\rsub{ 
From this analysis, we conclude that shift-out operations as implemented here have the potential for error rates below  the threshold of standard surface code implementations, but would ideally be reduced further  for scalable fault-tolerant operation. There are several options to reduce the ancilla measurement error. Using Stark shifts mediated by a longer lived excited state would reduce the error which is proportional to $\gamma^{2/3}$. For example 685 nm light could be used to couple to the $5d_{5/2}$ level which has a 1280 ns lifetime.  This straightforward upgrade would reduce the minimum error to about 0.07\%.  Alternatively one could use a calibrated light shift such that the targeted ancilla  sees a detuned rotation which cancels the population error in Eqs. (\ref{eq.c0c1})\cite{Xia2015}.
This can be made to work for both the $c_0$ and $c_1$ amplitudes since the oscillatory error frequencies are approximately integer multiples of each other at large detuning. 
Another alternative is to  perform targeted ancilla rotations  by combining global microwave operations with local $Z$-phase gates\cite{Graham2022} or by  use of optical Raman transitions at only the ancilla sites\cite{Knoernschild2010}. With this type of capability one could 
 apply the shelving pulses to all atoms, and then transfer only the ancilla sites from $\ket{3,0}\rightarrow\ket{4,1}$, which can then be mapped to $\ket{4,4}$ for the ancilla measurement.   }

\subsection{\rsub{ Ancilla Readout}}
\label{subsec.readout_improvements}

\rsub{ 
Improving the mid-circuit measurements to be compatible with fault-tolerant operation also requires improvements in the ancilla readout fidelity. We will discuss three main avenues to improve the readout fidelity. First, even though we have used a 0.7 NA lens to collect fluorescence, there are improvements in the readout optical train (optical train with lower aberrations and lower loss components) that would improve the collection efficiency. Such 
improvements should allow reaching up to 15\% efficiency (photoelectrons per scattered photon) using double sided 0.7 NA  imaging. }

\rsub{ 
Second, a major source of heating during the readout arises because the ground and excited states experience a different dipole trapping force strength. This difference causes heating when the atom scatters 852-nm light during the readout.  For the data presented in this manuscript, we have partially compensated this effect by detuning the readout light to dress the ancilla atom thereby reducing the heating from this effect \cite{Martinez-Dorantes2017}. However, a larger detuning slows the readout increasing the background noise during the readout and dephasing of the data qubits. Using a trap frequency that is magic for the ground to excited state transition or chopping the trap out of phase with the readout \cite{Kwon2017} should allow faster readout.
}

\rsub{ 
Finally, adding cooling via the 685-nm quadrupole transition will during the readout will prevent atoms from getting heating up as much during the readout process (see Appendix \ref{subsec.685cooling} below). Cooling will prevent atom loss during the readout, which will not only enable qubit resetting, but will also reduce readout noise. Atoms lost at different times during the readout will effectively be read out for different lengths of time. By preventing atom loss, all atoms see the same readout time and thus scatter photons during the entire readout duration. The equivalent readout times will result in a narrower distribution that enables bright and dark events to be classified with higher fidelity.  With these three experimental modifications we believe fault-tolerant mid-circuit measurements might be performed in less than 1 ms. }

\rsub{ 
Faster, higher fidelity readout will not only improve the mid-circuit measurement fidelity of the ancilla, but will also benefit the fidelity of the data qubits. A shorter readout time means that the data qubits do not need to be shelved in magnetically sensitive states as long. In addition, though the shelved states are 9.2 GHz detuned from resonance, the data qubits still see a spontaneous scattering rate of
\begin{equation}
    r_{\rm sc} = \frac{\gamma}{2} \sum_{f=2,3,4}\frac{I/I_{{\rm sat},f}}{1+4 (\Delta+\omega_q+\Delta_f)^2/\gamma^2 + I/I_{{\rm sat},f}},
    \label{eq.rsc}
\end{equation}
}

\rsub{ 
where $\gamma$ is the transition line width, $I$ is the intensity of light on the atoms, $I_{{\rm sat},f}$ is the saturation intensity for coupling to  excited state hyperfine level $f$, $\Delta$ is the detuning of the readout light from $f=5$, $\omega_q$ is the 9.2 GHz hyperfine splitting frequency, and $\Delta_f$ is the excited state hyperfine shift of level $f$ relative to $f=5$. Since $\omega_q\gg \gamma$ there is very weak saturation for the shelved atoms and for the purpose of estimating the rate we can simply add the inverse saturation intensities and average over the values found for atoms shelved in $\ket{3,-1}$ and $\ket{3,0}.$
Performing the hyperfine resolved angular momentum algebra we find the average scattering rate of the shelved qubits would be about 9\% less than that of an atom in the $\ket{4,4}$ stretched state at the same detuning value.}

\rsub{ 
Plugging in the readout parameters from Table \ref{Table.1}, we find a scattering rate of about $4.0~\rm s^{-1}$ for shelved data qubits, resulting in a fidelity error from scattering of  $1.6\%$ in a 4 ms readout. The calculated number of photons scattered by an ancilla atom in the bright state during the readout is approximately 9900. Referring to Fig. \ref{fig.Data} we see that 50 photoelectrons is sufficient for well resolved state measurements with an EMCCD camera. It should be possible with improved optics to push the overall detection efficiency with double sided imaging as shown in Fig. \ref{fig.1}, accounting for optical losses and camera quantum efficiency,  to 15\%. This implies a measurement can be made by scattering 330 photons. The corresponding ancilla measurement time and scattering error on the shelved data qubits would then be  130 $\mu\rm s$ and error of 
$5\times 10^{-4}$. These estimates support the feasibility of reaching performance compatible with quantum error correction with an improved imaging system. 
}

\section{Qubit resetting}
\label{sec.qubit_reset}

\subsection{Quadrupole recooling}
\label{subsec.685cooling}
In order for the mid-circuit measurement protocol to be compatible with multiple rounds of error correction, it is necessary to recool ancilla qubits after  the state measurement. In order to realize good two-qubit gate fidelity, the ancilla qubits should be recooled to under $10~\mu\rm K$. This recooling must be done without affecting the quantum state of shelved data qubits, and therefore requires a cooling protocol that does not couple to atoms in the lower $f=3$ hyperfine level. This requirement can be met by using narrow line cooling on the Cs quadrupole transition $6s_{1/2},f=4 \rightarrow 5d_{5/2},f=6$ at 685 nm. Preliminary measurements of laser cooling on this transition were reported in \cite{Carr2014t} and it has been proposed for implementation of a compact optical atomic clock\cite{Sharma2022}. 
\rsub{ It was demonstrated in \cite{Scott2023} that this transition can be used to cool atoms to $T=1.3~\mu\rm K$ and that the atoms could be imaged using 685 nm light without the use of a repupmer.}
Here we analyze an alternative version of 685-nm cooling using a single pair of counter-propagating cooling beams based on the Sisyphus cooling techniques proposed in \cite{Taieb1994, Ivanov2011} and demonstrated in Sr neutral atoms \cite{Covey2019a}.

\begin{figure}[!t]
\includegraphics[width=1.5in]{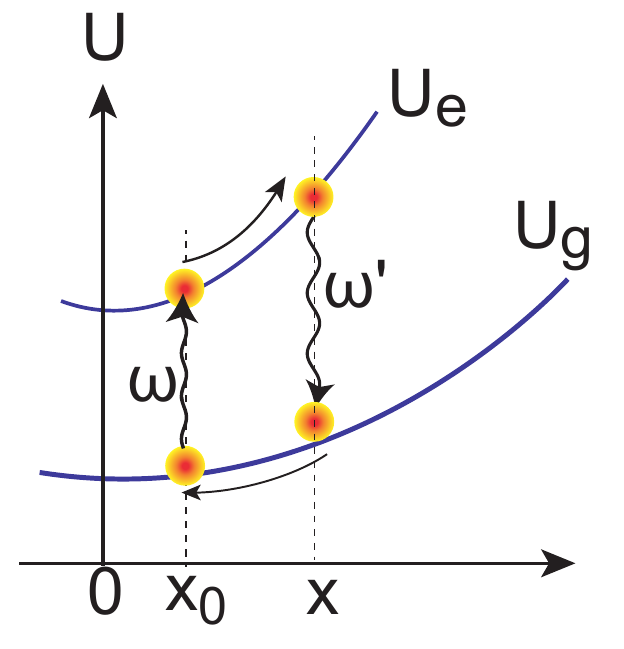}
\caption{
\label{SI-fig_Sisyphys}
Sisyphus cooling of a trapped atom. The atom absorbs a photon of frequency, $\omega$, at position, $x_0$ relative to the center of the trap. While the atom is in the excited state, it moves to a new location, $x$, where it emits a photon of frequency $\omega'$. If the excited state is more tightly confined than the ground state and the atoms moves further from the center of the trap, then $\omega < \omega'$ and the atom will experience cooling.
}
\end{figure}

\begin{figure}[!t]
\includegraphics[width=3.5in]{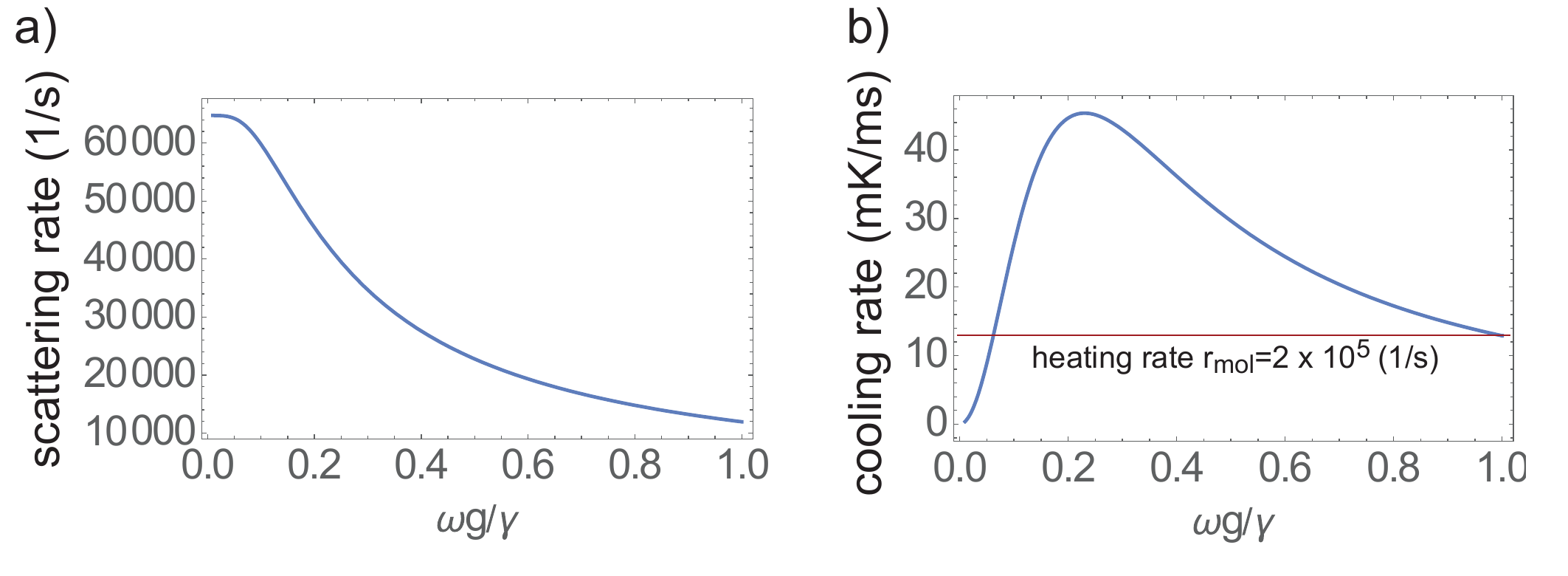}
\caption{
\label{SI-fig_Sisyphys_sim}
Sisyphus cooling rate with a decay rate $\gamma/2\pi=124~\rm kHz$, an excited state trap frequency $\omega_e=2\omega_g$ (where $\omega_g$ is the ground state trap frequency), a cooling beam  intensity $I/I_s=0.2$ (where $I_s$ is the saturation intensity of the transition), a cooling beam detuning from resonance at the center of the trap of $\Delta_0=0$, an atom temperature of $T=10~\mu\rm K$, and a trap depth of $U_{\rm trap}=500~\mu\rm K$ for a Gaussian beam trap with waist $w=1~\mu\rm m$. a) Shows the scattering rate of the cooling beams versus $\omega_g/\gamma$. b) Shows the cooling rate versus $\omega_g/\gamma$ relative to the transverse heating rate from 1D 852 nm molasses at the indicated $r_{\rm mol}$ scattering rate. 
}
\end{figure}

This technique uses a state-dependent trap depth and a narrow-line transition to enable cooling (see Fig. \ref{SI-fig_Sisyphys}). Specifically, a long-lived excited state is trapped more strongly than the ground state. The cooling lasers are tuned so that they excite atoms near the center of the trap up to the excited state. If the excited state has a long enough lifetime, then the atom has time to travel up the potential well before it decays back to the more weakly trapped ground state. Below we provide a semi-classical analysis of this cooling technique in 1D. In this cycle, the net energy change, $\Delta U$, of the atom is:
\begin{equation}
    \Delta U = -\left[ U_e(x_f) -U_e(x_0)\right]+\left[ U_g(x_f) -U_g(x_0)\right],
\end{equation}
where $U_g(x)$ ($U_e(x)$) is the trap potential of the ground (excited) state as a function of the initial (final) position of the atom position, $x_0$ ($x_f$), of the atom in the cooling cycle. Averaging over the time ($t$) the atom spends in the excited state and assuming harmonic potentials yields:
\begin{eqnarray}
\label{SI_eqn.deltaU}
     \delta U &=& -\frac{\int_{0}^{\infty} dt \Delta U e^{-\gamma t}}{\int_{0}^{\infty} dt e^{-\gamma t}} \nonumber \\
            &=& -\frac{\gamma m (\omega_e^2-\omega_g^2)}{2}\int^{\infty}_0 dt \left[ x^2(t) - x^2_0 \right] e^{- \gamma t},
\end{eqnarray}
where $\omega_g$ ($\omega_e$) is the trap frequency of the ground (excited) state, $m$ is the mass of the atom, and $\gamma$ is the decay rate of the excited state. An atom of temperature $T$ starting at $x_0$ at time $t=0$ will move in the excited state potential well with a trajectory:

\begin{equation}
     x_{\pm}= x_0 \cos(\omega_e t) \pm \sqrt{x^2_m - x^2_0} \sin(\omega_e t),
\end{equation}
where $x_m=\left( \frac{2 k_B T}{m \omega_e^2} \right)^{1/2}$ with $k_B$ representing Boltzmann's constant. Inserting $ x_{\pm}$ into eq. (\ref{SI_eqn.deltaU}) yields:
\begin{eqnarray}
    \delta U_\pm &=& -\frac{m(1-\omega_g^2/\omega_e^2)}{4(1+(\gamma/(2\omega_e))^2)} [\omega_e^2 (x_m^2-2 x_0^2) \nonumber \\
                &\pm& 2x_0\sqrt{x_m^2-x_0^2}\gamma\omega_e ]
\end{eqnarray}
Averaging the energy change then gives:
\begin{equation}
    \delta U = -\frac{m \omega_e^2 (1-\omega_g^2/\omega_e^2)}{4[1+\gamma^2/(4 \omega_e^2)]}(x_m^2-2x_0^2).
\end{equation}
To obtain a fast cooling rate, we need $\omega_e>\omega_g$ and $\gamma/\omega_e$ to be small. Also, an atom will only experience cooling when $x_0$ is close to the center of the trap.  In this technique, the cooling lasers are tuned so that the center of the trap has the maximum excitation rate; however, other positions in the trap have a finite excitation rate as well. To find the mean energy change, $\overline{\delta U}$, we need to average $\delta U$ over the excitation rate, $r(x)$:
\begin{equation}
    r(x)=\frac{\gamma}{2}\frac{I/I_s}{1+4 \left[ \Delta_0 + \Delta(x) \right]}.
\end{equation}
where $I$ is the intensity of the cooling laser, $I_s$ is the saturation intensity of the transition, $\Delta (x)$ is the position-dependent detuning of the of transition (the variation is due to the trap-induced Stark shift), and $\Delta_0$ is the detuning of the cooling lasers relative to the atomic resonance in the center of the trap. We need to weight the average over the atomic position distribution function, $\rho (x)$:
\begin{equation}
    \rho(x)=\frac{\gamma}{\sqrt{2 \pi \sigma}}e^{-x^2/(2\sigma^2)},
\end{equation}
where
\begin{equation}
    \sigma=\frac{w}{2}\sqrt{k_B T/U_{\rm trap}}.
\end{equation}
Here $T$ is the atom temperature, $w$ is the Gaussian beam waist of the trapping tweezer, and $U_{\rm trap}$ is the depth at the center of the trap.  The mean energy change per cooling cycle is then:
\begin{equation}
    \overline{\delta U}=\int_{-\infty}^{\infty} dx\,  \delta U(x) r(x) \rho (x).
\end{equation}
We have evaluated this integral numerically and found that we could achieve a cooling rate $\geq 40~\mu \rm K/ms$ for realistic trap parameters (see Fig. \ref{SI-fig_Sisyphys_sim}).  \rsub{ The attainable temperature depends on the linewidth of the excited state. Representative numerical simulations for Yb using the $^3P_1$ excited level show cooling to $8.5~\mu\rm K$ along each axis in a 3D trap\cite{Ivanov2011}. The Cs $5d_{5/2}$ level considered here has a 32\% narrower linewidth and we can expect correspondingly lower temperatures.}

This cooling technique has an advantage over using a six-beam red molasses in that we can configure the cooling beam pair to have a polarization that allows cycling from the $\ket{4,4}$ state without populating other $m_f$ levels. For the transition to be cyclic, the counter-propagating 685-nm cooling beam pair must be aligned such that their $k$-vector is perpendicular to the bias magnetic field and the beams have a polarization perpendicular to the quantization axis (e.g. if $B$  is along the $z$-axis, then the cooling beam $\bf k$-vectors are along $x$  and $-x$  and polarized along $y$ ). Quadrupole transition selection rules dictate that the possible $\Delta m_f$ of this transition can be -2, 0, or +2. We can suppress $\Delta m_f=-2$ and $0$ by providing a strong B-field that shifts those transitions out of resonance, so that the transition to $f'=6$, $m_{f'}=6$  is strongly preferred. The Land\'e g-factor for $5d_{5/2}$ $f'=6$ is $g_f'(f'=6,m_f'=6)=1/2$, so even the modest field of 10.2 G used in this paper will detune the $\Delta m_f=0$ and $\Delta m_f=-2$ transitions by 14.3 MHz and 28.6 MHz respectively, much greater than the 124 kHz linewidth of the transition. The $5d_{5/2}$ $f'=6$ $m_{f'}=6$ state will decay to $6s_{1/2}\ket{4,4}$  via the $6p_{3/2} \ket{5,5}$ state. The $\ket{4,4}$ state is the same state to which the mid-circuit readout procedure pumps ancilla atoms in the bright state. Therefore, we may directly recool atoms during the mid-circuit readout without affecting the dark state or the shelved qubits. In fact, since the $5d_{5/2}$ state decays to the ground state via the $6p_{3/2}$  state, each cooling cycle provides a photon that can be used to help detect the bright state.

\rsub{ 
Finally we note that the off-resonant scattering rate via the $5d_{5/2}, f'=5$ level is sufficiently low that ancilla atoms can be cooled without suffering Raman transitions to $f=3.$ Thus, even if Zeeman suppressed transitions with $\Delta m_f=-2, 0$ occur, the atom will still cycle back to $6s_{1/2}, f=4$ with very high probability. Ancilla atoms in the $f=4$ level can be repumped back to $f=4, m_f=4$ without disturbing shelved data qubits in $f=3$.  The cycling scattering rate via $5d_{5/2}, f'=6$ and the possibly non-cycling rate via $5d_{5/2}, f'=5$
are
\begin{eqnarray}
r_{\rm c}&=&\frac{\gamma}{2}\frac{I/I_{{\rm sat},6}}{1+4\Delta^2/\gamma^2+I/I_{{\rm sat},6}},\nonumber\\
r_{\rm nc}&=&\frac{\gamma}{2}\sum_{f=4,5}\frac{I/I_{{\rm sat},f}}{1+4(\Delta+\Delta_{{\rm hf},f})^2/\gamma^2)+I/I_{{\rm sat},f}}\nonumber.
\end{eqnarray}
It can be shown that the saturation intensity for the quadrupole transition from $6s_{1/2}, f=4$ to $5d_{5/2}, f'=4,5$ relative to the saturation intensity 
for the transition to $f'=6$ is 3.2 and 1.7 times larger, respectively, after averaging over Zeeman sublevels. Using $\Delta_{{\rm hf},5}=2\pi\times 127.4~\rm MHz $ and $\Delta_{{\rm hf},4}=2\pi\times 233.6~\rm MHz $ we find $r_{\rm nc}/r_{\rm c}=5.9\times 10^{-5}.$ This neglects the impact of branching ratios which will result in some of the photons scattered from 
$f'=4,5$ cycling back to $f=4.$ Thus the given value of $r_{\rm nc}/r_{\rm c}$ is a conservative upper limit for the non-cycling rate.  }

\rsub{ 
As discussed in Appendix \ref{sec.fault_tolerance}3 with improved optics it should be possible to make a well resolved ancilla measurement while scattering 330 photons. The corresponding transverse heating is $\sim 50 ~\mu\rm K$. To recool the ancillas we would conservatively need to scatter $\sim \sqrt{330}$ photons which implies a Raman probability from the 685 nm cooling light of 
approximately $0.001.$ }

\subsection{Quantum state reinitialization}
\label{sec.init}

After recooling, any ancilla qubits that were detected as bright during the mid-circuit readout need to be reset from $\ket{4,4}$ back to the initial state, $\ket{0}=\ket{3,0}$. This may be accomplished in a two step process: 1) Use microwave rotations to transfer from $\ket{4,4} \rightarrow \ket{4,1}$ using CORPSE pulses that are detuned from data qubits encoded in $m_f=0, -1$ states; 2) Perform site selective rotations on ancilla qubits, rotating $\ket{4,1} \rightarrow \ket{3,0}$. Step 1) of this resetting process would use the first four of the five CORPSE pulses which were used for state preparation described in the main text. Step 2) can be performed using methods described in either \cite{Graham2022} or site-selective Raman rotations \cite{Jones2007,Knoernschild2010}. If an ancilla qubit was determined to be in the dark state (initially in $\ket{1}=\ket{4,0})$, then it will be stored in  $\ket{3,0}$ during the mid-circuit measurement process and the final two-step global unshelving process will rotate the  ancilla 
back to $\ket{4,0}=\ket{1}$. Thus the ancilla will be restored to a known fiducial state depending on the measurement outcome.

\begin{figure*}[!t]
\center
\includegraphics[width=7in]{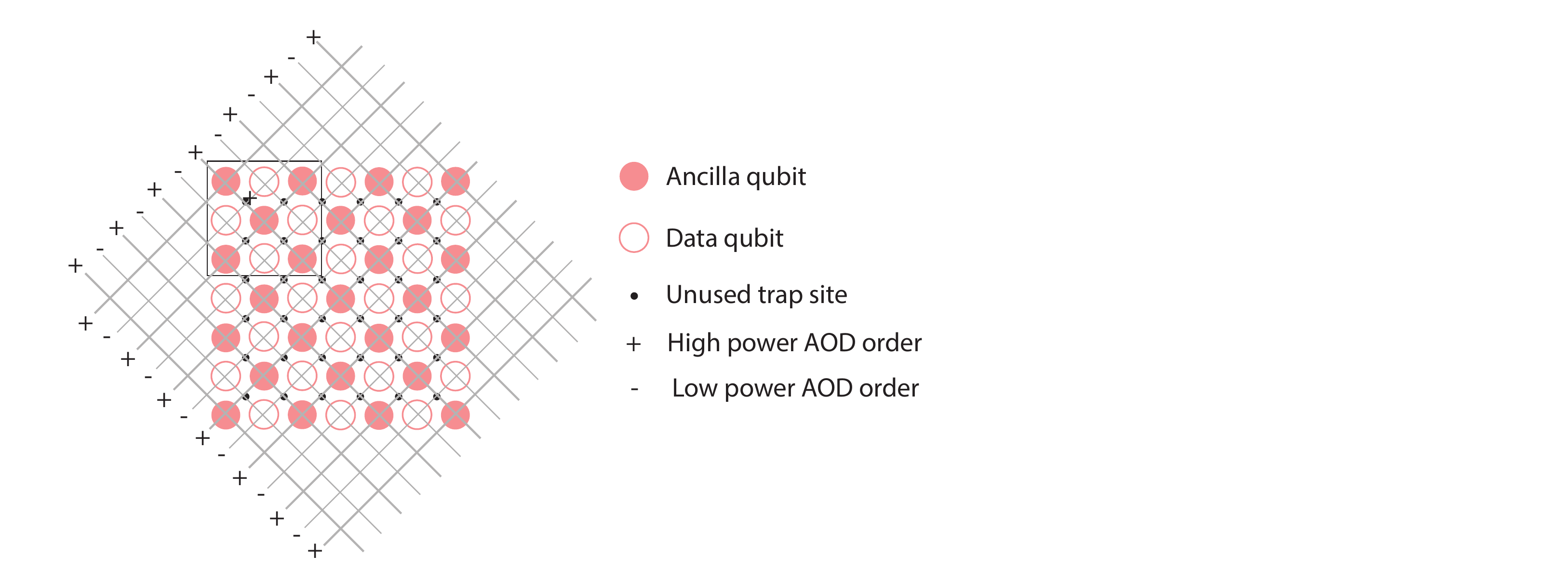}
\caption{
\label{SI_trap_ramping}
\rsub{ Scheme for scaling the trap ramping techniques to a large trap array with a layout compatible with the surface code. The black square  corresponds to the $3 \times 3$ array that was measured in this manuscript.  Initially all traps are uniform in power. In this configuration, the crossed acousto-optic deflectors are driven with frequency tones with amplitudes that have been adjusted such that the column and row  diffraction orders (oriented at $\pm$45 deg. in the figure) provide power $P_-$ to each order. During the mid-circuit readout, the acousto-optic deflector tones are adjusted such that alternate column and row orders are increased in power to $P_+$. This reduces the diffraction into the $-$ orders, so the optical power into the deflectors is simultaneously increased (using an upstream power modulator) to keep the power $P_-$ constant.  This sequence increases the ancilla trap depths without changing the data qubit trap depths. The unused trap sites can be kept empty or be used to store replacement atoms that are shelved in dark states that are not measured and are not coupled to Rydberg levels, so as not to disturb two-qubit gate operations.  }
}
\end{figure*}

\section{\rsub{ Scalability}}
\label{Scalability}

\rsub{ 
In this manuscript, we have demonstrated mid-circuit measurement in a small array with a single ancilla and four data qubits. In this section we explore paths to scaling these techniques to a larger atom array. The two main adaptations required for performing mid-circuit measurements in a large atom array involve trap ramping and single qubit rotations. Our mid-circuit measurement demonstration used trap ramping to increase the trap depth of the ancilla qubit to allow more photons to be scattered before the atom was heated out of the trap. Since our trap arrays were made using two orthogonally-oriented acousto-optic deflectors (AODs), we were limited in how we could adjust the trap depths and could only adjust the depth of an entire row and/or column rather than an individual site. To ramp up the ancilla trap (the central site of the $3 \times 3$ array) without affecting the data qubits, we decreased/increased the depth of a row and column in an alternating fashion. The resulting power adjustment increases the depth of the ancilla trap without affecting the power of its four nearest neighbors. However, the four corner traps are reduced in depth making them unsuitable for data or ancilla qubits, though they might still be used as an atom reservoir for atomic rearrangement. }

\rsub{ 
A scaled-up extension of this type of technique to a large array can be obtained using the geometry and ramping procdeure  described in Fig. \ref{SI_trap_ramping}. This results in interleaved arrays of data and deep ancilla sites compatible with the surface code architecture\cite{Fowler2012}.  If ancilla measurements with improved optics are used to limit the number of scattered photons and atom heating (see discussion in Appendix \ref{sec.fault_tolerance}3) no trap ramping would be required. In such a case spatial light modulator\cite{DKim2019} or other techniques\cite{Huft2022} for making larger 2D arrays could be used. }

\rsub{ 
Consideration is also needed in scaling the single qubit gates. The local shift-out operations, or other techniques described in Appendix \ref{sec.fault_tolerance}2, can be performed in parallel on multiple ancilla sites using AODs driven by multiple tones. Since these operations involve very large optical detunings, or two-photon Raman operations, they are insensitive to the frequency shifts caused by the AODs. Spatial light modulators could also be used for simultaneous addressing of multiple ancilla sites\cite{Graham2023}. 
}
\end{document}